\documentclass{article} % Adjust as needed
\usepackage{booktabs}
\usepackage[margin=0.8in]{geometry} % Sets 1-inch margins on all sides
\usepackage{graphicx} % Required for inserting images
\usepackage{xcolor}
\usepackage{hyperref}
\usepackage{tabularray}
\usepackage[
  backend=biber,
  style=apa,
  sorting=nyt,
  maxcitenames=2,
]{biblatex}
\usepackage{amsmath}
\usepackage{comment}
\usepackage{booktabs}
\usepackage{adjustbox}
\usepackage{makecell}
\usepackage{authblk}  % For author and affiliation blocks
\usepackage{pdflscape}
\usepackage{amsfonts}
\usepackage{tabularx}
\usepackage{array}
\usepackage{ragged2e}
\usepackage{longtable}
\usepackage{subcaption}
\usepackage{tcolorbox}
\tcbuselibrary{listings, breakable}
\newcommand{\conv}{\tau}
\newcommand{\pers}{\rho} 
\newcommand{\expo}{\eta}

\let\cite\parencite

\title{Conviction, persuasiveness, and cross-cutting exposure: a Bayesian model of voting behaviour in online political debates}

\author[1,2,*]{Elena Candellone}

\affil[1]{Department of Methodology, Statistics \& Data Science, Utrecht University, Utrecht, Netherlands}
\affil[2]{Centre for Complex Systems Studies, Utrecht University, Utrecht, Netherlands}
\affil[*]{Corresponding author. Email: \href{e.candellone@uu.nl}{e.candellone@uu.nl}}

\date{\today}

\begin{document}
%TC:ignore
\maketitle
\begin{abstract}
Online debate platforms offer a window into whether exposure to diverse crowds changes how people reason: they capture political preferences, the perceived quality of arguments, and the ideological diversity of the audience. Drawing on theories of deliberative democracy and cross-cutting exposure, I develop a Bayesian logistic regression model that decomposes voting behaviour in online debates into three components: conviction (agreement between a user's prior beliefs and a debate's stance), persuasiveness (perceived argument quality), and cross-cutting exposure (the ideological diversity of a debate's prior voters). I apply this framework to the DDO dataset: approximately 341k votes across 78k debates on 48 socio-political topics. As the platform provides no predefined topic labels, I infer topic and stance from debate text using large language models. I test whether a more ideologically diverse prior audience weakens a voter's conviction, a central claim from the cross-cutting exposure literature. Persuasiveness is the strongest and most consistent predictor of vote choice, never outweighed by conviction, though conviction remains a credible predictor on several topics tied to personal freedoms. Cross-cutting exposure moderates conviction on 4 of 47 topics, though these results are sensitive to topic classification and sparse coverage, with implications for the design of deliberative platforms.

\bigskip

\noindent\textbf{Keywords: online political debates; ideal point models; Bayesian inference; deliberative democracy; cross-cutting exposure.}

\end{abstract}
%TC:endignore

\section{Introduction}

Online platforms have fundamentally changed the way humans discuss political content and form their opinions about societal issues.  They have created new opportunities to connect with virtually everyone with an internet connection around the globe and to consume and produce content across a wide variety of digital platforms, ranging from instant messaging to online debates. It has been an unprecedented avenue for both the improvement and worsening of Western democratic institutions, creating a space for sharing and shaping political views~\cite{adamic2005political, flamino2023political, tornberg2022digital}.  Online debate platforms, in particular, aim to simulate an idealized, Athenian-democracy-reminiscent scenario: users can debate and deliberate on an online forum rather than in a Greek \emph{agora}. They aspire to simulate deliberative-democratic institutions such as citizens' assemblies and mini-publics, in which structured exchange among ordinary citizens is meant to produce considered collective judgments, and not simply to aggregate pre-existing preferences~\cite{chambers2025deliberation}. This democratic ideal is criticized by radical-democracy theorists, such as Mouffe, who reject the deliberative aim of reaching consensus, arguing that liberal democracy has a paradoxical nature, torn between individual liberty and popular sovereignty, and that political conflict is not solvable by a better argument~\cite{mouffe2000democratic}. Whether online debate platforms are closer to the deliberative ideal or to its radical-democratic critique, i.e., whether votes track the reasoned weighing of arguments, or simply reproduce the composition of prior beliefs, is the empirical question this paper addresses. In fact, from a research perspective, these platforms are invaluable tools; more traditional survey settings often lack information on social influence, making it hard to know whether a respondent's answer has been conditioned by someone else. On the other side, experimental settings may not always have information on explicit opinions; in that case, we rely on observed behaviours rather than self-reported attitudes. Online debate platforms contain both the opinions expressed by individuals and the social voting context in which those opinions are formed. This makes them a uniquely transparent window into the mechanisms driving opinion formation on socio-political topics.

When a user votes in an online debate, at least three distinct inputs are available. The first is their own prior belief on the topic at hand; the second is the argumentation presented by the two debaters; the third is the ideological composition of the audience that has already voted. Debate.org was unusual in recording all three: users self-report their stance on 48 political topics; they choose which debater made the more convincing arguments; and their votes are timestamped within a visible sequence, making the temporal order of votes observable. Separating these inputs matters because they carry different implications for the character of deliberation on the platform. If prior beliefs mainly predict votes, the debate does not help users update pre-formed positions; if they track perceived argument quality, the platform is functioning as a space for deliberation~\cite{chambers2025deliberation}; and if the diversity of the audience that came before shapes how strongly a voter relies on their own prior belief, this is evidence for or against the claim that exposure to diverse crowds changes how people reason.

The first mechanism draws on the moral conviction literature~\cite{skitka2010psychology,skitka2021psychology,maxey2026moralization}. A moral conviction is believing that a certain opinion is held because of core values. Beliefs can become moral convictions through social and cultural processes, and once they do, they resist pressure from peers~\cite{skitka2021psychology,aramovich2012opposing}. Stereotypes and affective polarization can emerge from social interaction and coherence-seeking alone, without any real correlation between group membership and the stereotyped preference~\cite{seckin2026emergence}. The model defined here operationalizes conviction by estimating independent effects of conviction from persuasiveness and cross-cutting exposure: motivated-reasoning and Bayesian literature on belief updating treat it as a joint product of prior commitment and evidence~\cite{Kunda1990,Priniski2022,Rigoli19012026,maher1993betting, bartels2002beyond}, so conviction can resist peer pressure while remaining open to argument, consistent with evidence that identity-relevant attitudes such as vaccine hesitancy shift under evidence-based interventions~\cite{Horne2015, Powell2023}. Issues of personal autonomy and lifestyle are a canonical case of \textit{morality policy}, conflicts over first principles and personal values~\cite{mooney2001clash}, and are accordingly expected to show the largest conviction effects here, though not necessarily effects that outweigh persuasiveness, given the platform design.

The second mechanism concerns persuasiveness: whether a vote tracks the debater judged more convincing, independent of prior beliefs. In the persuasion literature, the elaboration likelihood model distinguishes a \textit{central route}, in which recipients respond to actual argument quality, from a \textit{peripheral route}, in which surface cues such as source credibility or style do the persuasive work instead~\cite{petty1986elaboration}. Work on related web-argumentation platforms supports this distinction: convincingness on \textit{createdebate.com} and \textit{convinceme.net} is modeled as a qualitative property of arguments and can be explained in terms of their strengths and flaws~\cite{habernal2016convincing}, while persuasion in good-faith online discussion more broadly is predicted by an argument's interaction dynamics and language~\cite{tan2016winning}. The deliberative quality of such debates, moreover, varies systematically across platform types, with moderated forums yielding more rational and respectful discourse than social media pages~\cite{esau2021arenas}. Persuasiveness dominating conviction would be the empirical signature of central-route processing, i.e., deliberation closer to the ideal described by Chambers~\cite{chambers2025deliberation}.

The third mechanism draws from the cross-cutting political exposure literature~\cite{mutz2006hearing}, which studies how the ideological heterogeneity of the social environment shapes political attitudes. Exposure to disagreement typically moderates reliance on prior conviction: people embedded in more heterogeneous environments express more ambivalence and are less likely to act on their priors alone, while those in homogeneous environments show more confident, more polarized expression~\cite{sunstein2009going}.  This is not a universal finding, however; some studies find entrenchment or reactance instead~\cite{bail2018exposure}, making the moderation direction an empirical question. Debate.org's visible voting sequence makes it possible to construct, for each voter, an ideological-diversity measure of the audience that voted before them, and test directly whether it moderates the conviction-vote association.

Despite their importance, most direct evidence on how people update judgments in the presence of others comes from controlled experiments on generic conformity and herding, e.g., artificial markets, randomized rating manipulations, and estimation tasks~\cite{lorenz2011social,muchnik2013social,salganik2006experimental}, rather than from real-world political settings that separate conviction, persuasiveness, and cross-cutting exposure as distinct, jointly operating mechanisms. To the best of my knowledge, no prior work has unified these three mechanisms within a single framework, or tested their relative importance across topics in a non-experimental, real-world political setting.

A related but distinct body of work addresses voting behaviour in observational data: part of the political science literature studies real political voting by estimating the latent ideology of legislators from their votes on bills~\cite{poole1985spatial, clinton2004statistical}. Ideal point models, reminiscent of item-response models from psychometrics~\cite{baker2004item}, infer a latent ideological dimension from voting patterns under the assumption that legislators with similar ideologies vote similarly. These models have been extended to social media applications by Barberá~\cite{barbera2015birds, barbera2015tweeting}, who applied them to Twitter/X data to jointly infer the latent ideology of users and political elites based on their reposting behaviour. This is a powerful framework to recover latent ideological positions from interactional data, but it does not seek to decompose the drivers of voting behaviour: it does not model mutual influence among voters, nor does it ask whether a given vote reflects prior convictions or other voting mechanisms. These are precisely the questions this paper addresses.

\begin{figure}
    \centering
    \includegraphics[width=\linewidth]{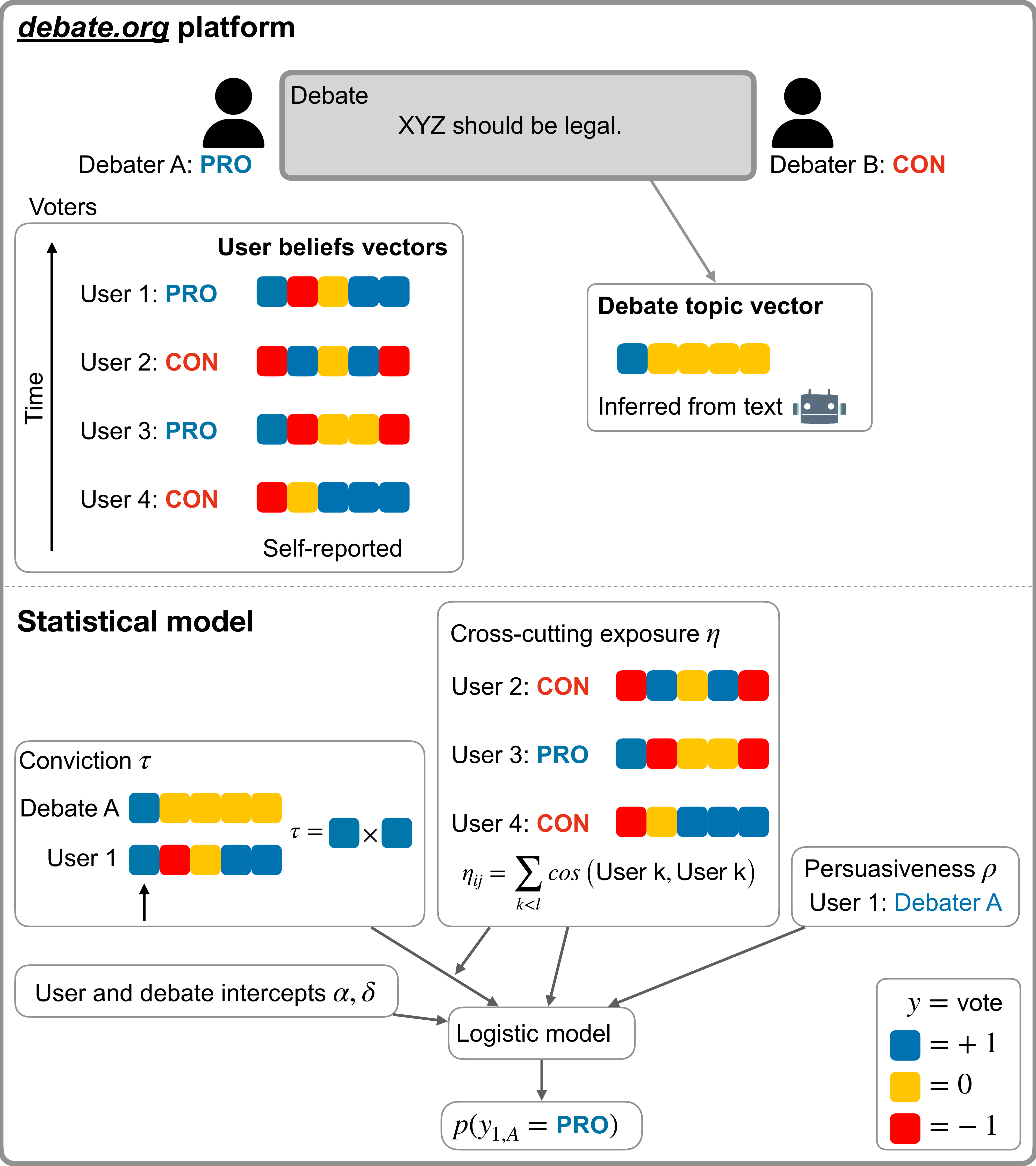}
 \caption{\textbf{Platform structure and model description.} Each debate (e.g. ``XYZ should be legal'') has one \textbf{PRO} and one \textbf{CON} debater. Users can also vote \textbf{PRO}/\textbf{CON}. Each user has a belief vector of length K, with their self-reported stance on each of the K topics. Each debate has a topic vector of the same length, inferred from the debate text (see Section~\ref{par:topic-stance-inference}). I use these vectors to compute the conviction and cross-cutting exposure terms of the logistic model that predicts whether a user would vote \textbf{PRO}/\textbf{CON}.}
    \label{fig:model}
\end{figure}

To do so, I develop a Bayesian hierarchical logistic regression model, inspired by ideal point models, to jointly estimate these mechanisms across a broad range of socio-political topics. I apply this framework to the \textit{DDO} dataset~\cite{durmus-cardie-2018-exploring, durmus-cardie-2019-corpus}, collected from the online debating platform \textit{debate.org} (see Fig.~\ref{fig:model}). On this platform, users vote \textbf{PRO} or \textbf{CON} on structured debates and can view the running vote tally before casting their own vote, making the ideological composition of prior voters observable and reconstructable from the data. Three components are operationalised as predictors of voting behaviour: \textit{conviction}, capturing the similarity between the inferred debate topic and the user's self-reported prior beliefs; \textit{persuasiveness}, recording which debater the user judged to have argued more convincingly; and \textit{cross-cutting exposure}, capturing the ideological diversity of the debate's prior voters. As cross-cutting exposure theory concerns the moderation of conviction, I model it as an interaction with the conviction term, alongside its own main effect. This framework makes it possible to address the paper's central research questions empirically:

\textit{\textbf{RQ1:} Which is the stronger predictor of voting: conviction or persuasiveness?}

\textit{\textbf{RQ2:} Does cross-cutting exposure change the effect of conviction?}

I find that persuasiveness is the strongest and most consistent predictor of vote choice: 42 of 47 topics show a strong, reliable association with perceived persuasiveness (coefficients credibly distinguishable from the null), versus 26 of 47 for conviction, and conviction never credibly outweighs persuasiveness on any topic. Sixteen topics are \textit{persuasiveness-only} (strong persuasiveness, no comparably strong conviction); eighteen are \textit{persuasiveness-driven} (both coefficients strong, persuasiveness is the larger of the two); and conviction contributes a strong, comparably sized effect on eight further topics (\emph{jointly-driven}, no clear dominance either way), several of which are tied to personal freedom and lifestyle choices such as legalised prostitution and medical marijuana. Cross-cutting exposure credibly moderates the effect of conviction on 4 of 47 topics: it weakens conviction's pull, as predicted by Mutz's cross-cutting exposure theory, on 3 topics (animal rights, death penalty, progressive tax), and strengthens it, consistent with reactance/entrenchment accounts, on 1 (smoking ban), while the remaining 43 topics show no credible moderation. Section~\ref {sec:conclusion} discusses these findings and their implications for the stability of online political discourse.

\clearpage

\section{Data}
\subsection{DDO dataset}\label{subsec:data}
The DDO dataset was collected by Durmus and Cardie~\cite{durmus-cardie-2019-corpus,durmus-cardie-2018-exploring} from the online debating platform \textit{debate.org}, active between 2007 and 2022. The platform enabled users to engage in structured debates by taking opposing sides (\textbf{PRO}/\textbf{CON}) on a given topic across multiple rounds, after which third-party users could vote \textbf{PRO}, \textbf{CON}, or \textbf{TIE} to determine the winner.

The dataset comprises 78,376 debates conducted between October 2007 and November 2017, involving 46,475 unique voters and a total of 340,692 votes. It has been studied from the perspective of beliefs research~\cite{lee2025semantic, lee2026llms}, reflecting the richness of its user-level data on opinions and demographics. In addition to their voting activity, users could self-report their stance on 48 predefined political and social issues (hereafter \emph{topics})~\footnote{The 48 topics are: Abortion, Affirmative Action, Animal Rights, Barack Obama, Border Fence, Capitalism, Civil Unions, Death Penalty, Drug Legalization, Electoral College, Environmental Protection, Estate Tax, European Union, Euthanasia, Federal Reserve, Flat Tax, Free Trade, Gay Marriage, Global Warming Exists, Globalization, Gold Standard, Gun Rights, Homeschooling, Internet Censorship, Iran-Iraq War, Labor Union, Legalized Prostitution, Medicaid \& Medicare, Medical Marijuana, Military Intervention, Minimum Wage, National Health Care, National Retail Sales Tax, Occupy Movement, Progressive Tax, Racial Profiling, Redistribution, Smoking Ban, Social Programs, Social Security, Socialism, Stimulus Spending, Term Limits, Torture, United Nations, War in Afghanistan, War on Terror, Welfare.}, along with other demographics, such as education, ethnicity, gender, age, religion, political party and ideology, and favourite U.S. president.

I encode each user $i$ by a belief vector $u_i \in \{-1,0,+1\}^K$, where $K=48$ is the number of topics and $u_{i,t}$ indicates the user's stance on topic $t$: $+1$ (support), $-1$ (oppose), or $0$ (no reported opinion). I retain only users who expressed at least one prior belief, resulting in a sample of 15,840 users. Appendix~\ref{app:user-demographics} describes the user demographics and users' prior beliefs. The user base is mainly young, with a majority of white, male users.

For interpretability and to identify patterns across thematically related topics, topics are grouped into nine macro-categories (hereafter, \textit{issues}) following the ANES classification~\cite{anesGuide}, as shown in Table~\ref{tab:topic-issues}.  These groupings are used both to organise the visual presentation of results and to examine whether the relative contribution of conviction and persuasiveness varies systematically across issue domains.

\begin{longtblr}[
  caption = {Grouping of \textit{debate.org} topics into ANES-based~\cite{anesGuide} issues.},
  label = {tab:topic-issues},
]{
  width = \linewidth,
  colspec = {Q[450]Q[750]},
  row{1} = {font=\bfseries},
  hline{1-2,11} = {-}{},
}
Issue                           & Topics                                                                                                                                                                                                 \\
Social welfare, healthcare, and education & homeschooling, medicaid  medicare, national health care, social security, social programs, welfare                                                                                                     \\
Economy, budget, and taxes                & capitalism, estate tax, federal reserve, flat tax, gold standard, labor union, minimum wage, national retail sales tax, occupy movement, progressive tax, redistribution, socialism, stimulus spending \\
Social and cultural issues                & abortion, civil unions, drug legalization, euthanasia, gay marriage, internet censorship, legalized prostitution, medical marijuana, smoking ban                                                       \\
Guns and crime                            & death penalty, gun rights, torture                                                                                                                                                                     \\
Military and foreign policy               & european union, iran-iraq war, military intervention, united nations, war in afghanistan, war on terror                                                                                                \\
Immigration, trade, and globalization     & border fence, free trade, globalization                                                                                                                                                                \\
Race                                      & affirmative action, racial profiling                                                                                                                                                                   \\
Environment                               & animal rights, environmental protection, global warming exists                                                                                                                                         \\
Government and politics                   & barack obama, electoral college, term limits                                                                                                                                                           
\end{longtblr}

\subsection{Measures}\label{subsec:measures}

In this section, I define the four quantities used to characterize each vote: conviction ($\conv$), persuasiveness ($\pers$), cross-cutting exposure ($\expo$), and the platform-collected user-level demographic covariates. Conviction, persuasiveness, and cross-cutting exposure enter the regression model directly, as described in Section~\ref{sec:methods}; the covariates do not, for reasons discussed at the end of this section. Conviction is built from each debate's inferred topic vector $d_j$, so I describe its construction first.

\paragraph{Inferring debate topic and stance.}\label{par:topic-stance-inference}
The platform does not label debates with one of the 48 topics. Therefore, to construct a topic representation comparable to user belief vectors, I infer for each debate $j$ a topic index $t_j \in \{1,\dots, K\}$ and an associated stance.

For this task, I employ GPT-4o-mini with zero temperature to ensure a deterministic output. The model takes the debate title and full text as input for topic inference, and the title alone for stance classification; for prompt engineering, I followed the methodological guide~\cite{qixiang2026}. Using the title alone for stance is a deliberate choice: the full debate text often contains arguments on both sides, which can confound stance detection, whereas the title typically frames the debate from a single directional perspective (e.g., ``X should be abolished'').
As a robustness check, I compare results against a pipeline combining Sentence-BERT~\cite{reimers2019sentence} for topic detection and BART-large-MNLI~\footnote{\url{https://huggingface.co/facebook/bart-large-mnli}} for stance detection. Results are relatively stable across both classifications, as reported in Appendix~\ref{app:debate-classification}.

Each debate is assigned a single topic and a stance (\textbf{PRO}, \textbf{CON}, or \textbf{NEUTRAL}). Debates classified as ``other'' topic (over half of cases) are excluded from the analysis. For the remaining debates, I construct a topic vector $d_j \in \{-1,0,+1\}^K$ with a single non-zero entry at position $t_j$, equal to $+1$ for \textbf{PRO} or \textbf{NEUTRAL} stances and $-1$ for \textbf{CON} stances. \textbf{NEUTRAL} stances are grouped with \textbf{PRO}, as they do not convey opposition but rather descriptive or exploratory framing (e.g., ``An overview of the debate surrounding~X''). In such cases, the debate presents the topic without taking a side against it, making a positive or neutral alignment with the topic the more faithful representation. Grouping \textbf{NEUTRAL} with \textbf{CON} would incorrectly imply opposition where none is expressed; treating them as a separate category would produce an all-zeros debate vector, which would zero out the conviction term $\conv_{ij}$ for all voters on that debate, effectively discarding potentially informative observations from the analysis.

\paragraph{Conviction.}
The \textit{conviction} term $\conv_{ij} \in [-1,1]$ measures the ideological proximity between user $i$'s prior beliefs and debate $j$'s inferred stance.
Because $d_j$ has a single non-zero entry at position $t_j$, equal to +1 for \textbf{PRO} debates and -1 for \textbf{CON} debates, $\conv_{ij}$ reduces to a normalized version of the user's prior belief on topic $t_j$ ($u_{i,t_j}$), signed by the debate stance on that topic ($d_{j,t_j}$):
\begin{equation}
    \conv_{ij} = \frac{u_{i,t_j} \, d_{j,t_j}}{\|u_i\|}.
    \label{eq:tau}
\end{equation}
In practice, $\conv_{ij}>0$ when the user's prior belief is aligned with the debate stance, $\conv_{ij}<0$ when they are opposed, and $\conv_{ij} = 0$ when the user has no reported prior belief on that topic.

\paragraph{Cross-cutting exposure.}
The \textit{cross-cutting exposure} term $\expo_{ij} \in [-1,1]$ measures the ideological diversity of prior voters on debate $j$, via the negative mean pairwise cosine similarity among their belief vectors:
\begin{equation}\label{eq:D}
\expo_{ij} =
\begin{cases}
\displaystyle -\frac{2}{|V_{j,i}|(|V_{j,i}|-1)}\sum_{\substack{k,l\in V_{j,i}\\k<l}}\cos(u_k,u_l) & \text{if } |V_{j,i}|>1,\\[1.5ex]
0 & \text{if } |V_{j,i}|\leq 1,
\end{cases}
\end{equation}
where $V_{j,i} = \{k\neq i: TS_{kj}\leq TS_{ij}\}$ is the set of voters to debate $j$ before user $i$. $\expo_{ij}$ is near $-1$ when prior voters are ideologically homogeneous, near $0$ when their beliefs are uncorrelated, and near $+1$ when they are systematically opposed. Unlike conviction, $\expo_{ij}$ does not use user $i$'s own belief vector, so it is not confounded with $\conv_{ij}$ by construction. It has one limitation (Appendix~\ref{app:diversity-simulation}): as audience size grows, same-camp and cross-camp pairs offset in the average, so a large polarized two-camp audience converges toward the same value as an ideologically uncorrelated one. This bears little on the moderation hypothesis, since cross-cutting exposure theory~\cite{mutz2006hearing} concerns exposure to disagreement per se, but it matters for the entrenchment reading, which implicates a cohesive opposing camp rather than diffuse heterogeneity~\cite{bail2018exposure}.

\paragraph{Persuasiveness.}
The \textit{persuasiveness} term $\pers_{ij} \in \{-1, 0, +1\}$ records which debater user $i$ judged to have made the more convincing arguments, signed by that debater's position: +1 when the user credited the \textbf{PRO} debater and -1 when they credited the \textbf{CON} debater, 0 if they did not believe any of the debaters was convincing. Unlike the other two predictors, this term is not constructed from the user or debater vectors, and so does not share statistical dependence.

\paragraph{Demographic covariates.}
Users could also self-report demographic information, including education, ethnicity, gender, age, religion, political party, political ideology, and favourite U.S. president (Appendix~\ref{app:user-demographics} reports their distribution). These covariates are \emph{not} included as predictors in the regression model (Eq.~\ref{eq:model}), for two reasons. First, the paper's two research questions concern the relative weight of conviction, persuasiveness, and cross-cutting exposure within a vote, a within-user comparison already absorbed by the user-level random intercept $\alpha_i$ regardless of its demographic source. Second, self-reporting was voluntary and non-response is substantial and uneven across dimensions (Appendix~\ref{app:user-demographics}), which would complicate their use as covariates without a principled missing-data treatment. I therefore use them only to characterize the composition of the user base, not as model predictors.
\clearpage
\section{Methods}\label{sec:methods}
\subsection{Bayesian hierarchical logistic-regression model}

\subsubsection{Assumptions}
My modelling choices are based on four assumptions. First, each observed vote is treated as a binary outcome, \textbf{PRO} or \textbf{CON}, conditional on participation; abstentions and ties are excluded from the analysis.

Second, user preferences (prior beliefs) and debate content (inferred topics) are represented in a common $K$-dimensional space. Ideological proximity between a user and a debate is captured by a normalized alignment measure based on their respective representations.

Third, the \textit{cross-cutting exposure} term represents the ideological diversity of users who cast their vote on the same debate before user $i$. Unlike the other two terms, this term is not treated as a direct, competing predictor of the vote; based on cross-cutting exposure theory~\cite{mutz2006hearing}, I test it as a moderator of the conviction term, alongside its own main effect.

Fourth, unobserved heterogeneity at both the user and debate levels is represented via additive random intercepts, $\alpha_i$ and $\delta_j$, described further below.

\subsubsection{Statistical model}
I model voting behaviour with a \textbf{Bayesian hierarchical logistic-regression model}. The model specification is inspired by ideal point models widely used in political science to study legislative voting~\cite{jackman2001multidimensional,clinton2004statistical,poole1985spatial}, and by their adaptation to social media settings~\cite{barbera2015birds,barbera2015tweeting}. I depart from Barberá's framework in three aspects. First, whereas Barberá models the influence of a political elite on a user's majority, I model the association between a user's vote and the ideological proximity of prior beliefs and the debate stance. Second, I incorporate a direct measure of argument persuasiveness. Third, rather than treating audience composition as a predictor competing with conviction, I model it primarily as a moderator, following the theoretical claim that cross-cutting exposure changes how strongly people rely on prior beliefs~\cite{mutz2006hearing}.

Formally, let $i \in \{1,\dots, N\}$ index users and $j \in \{1,\dots, M\}$ index debates, and let $t_j \in \{1,\dots,K\}$ denote the topic associated with debate $j$. Conditional on user $i$ casting a vote on debate $j$, let $y_{ij} \in \{0,1\}$ denote the observed outcome (i.e. encoded from the answer to the question ``Agree with after the debate''), where $1$ corresponds to \textbf{PRO} and $0$ to \textbf{CON}. The probability of a \textbf{PRO} vote is modeled as:
\begin{equation}\label{eq:model}
P(y_{ij}=1\mid\cdot) = \sigma\Big(
\underbrace{\alpha_i}_{\text{user int.}} + \underbrace{\delta_j}_{\text{debate int.}}
+ \underbrace{\beta_{\conv}[t_j]\,\conv_{ij}}_{\text{conviction}}
+ \underbrace{\beta_{\pers}[t_j]\,\pers_{ij}}_{\text{persuasiveness}}
+ \underbrace{\beta_{\expo}[t_j]\,\expo_{ij}}_{\text{cross-cutting exposure}}
+ \underbrace{\beta_{\conv \expo}[t_j]\,(\conv_{ij}\times \expo_{ij})}_{\text{conviction} \times \text{exposure}}
\Big),
\end{equation}

where $\sigma(\cdot)$ denotes the logistic function. Eq.~\ref{eq:model} is stated for the case in which there is a prior audience; for votes with none (a debate's first vote, or debaters voting on their own debate), the $\beta_\expo[t_j]\expo_{ij} + \beta_{\conv \expo}[t_j](\conv_{ij}\expo_{ij})$ terms are replaced by a single global intercept $\gamma_{\text{no-prior}}\sim\mathcal{N}(0,1)$ (Appendix~\ref{app:no-prior}).

The model has two kinds of terms. $\alpha_i$ and $\delta_j$ are \textbf{random intercepts}: $\alpha_i$ captures user $i$'s persistent, topic-independent tendency to vote \textbf{PRO} regardless of debate content, and $\delta_j$ captures debate $j$'s persistent tendency to attract \textbf{PRO} votes regardless of who is voting (e.g., its overall controversiality or framing). Together they absorb unobserved user- and debate-level heterogeneity not explained by the three substantive predictors.

$\beta_\conv[t]$, $\beta_\pers[t]$, $\beta_\expo[t]$, and $\beta_{\conv\expo}[t]$ are \textbf{topic-specific random slopes}. Substantively, $\beta_\conv[t]$ is the association between a user's prior-belief alignment and their vote on debates about topic $t$; $\beta_\pers[t]$ is the association between the perceived persuasiveness of a debater's arguments and the vote; $\beta_\expo[t]$ is the main-effect association between prior-audience diversity and the vote; and $\beta_{\conv\expo}[t]$ is the topic-specific interaction coefficient, capturing whether prior-audience diversity strengthens or weakens the pull of conviction on that topic. $\beta_{\conv\expo}[t] < 0$ indicates that greater prior-audience diversity is associated with a weaker conviction effect, consistent with Mutz's hypothesis; $\beta_{\conv\expo}[t] > 0$ indicates the opposite, consistent with the reactance or entrenchment hypothesis~\cite{bail2018exposure}. Allowing these four coefficients to vary by topic is what lets the model address the paper's research questions: whether the relative importance of conviction and persuasiveness differs across socio-political topics (RQ1), and whether cross-cutting exposure changes the effect of conviction (RQ2).

All coefficients in this observational model should be interpreted as conditional associations rather than causal effects. For $\beta_\expo[t]$ and $\beta_{\conv\expo}[t]$ in particular, an association between a debate's prior audience and a user's vote could reflect a genuine effect of that audience, or users self-selecting into debates whose audience already resembles them. I address this concern, and the placebo test it motivates, in Appendix~\ref{app:diversity-placebo}.

\subsection{Bayesian parameter estimation}
The model is estimated in a Bayesian framework using Stan \cite{stan}. Because $\conv$, $\pers$, and $\expo$ differ substantially in raw scale (SDs of approx. $0.18$, $0.61$, and $0.11$, respectively), all three are $z$-scored before estimation, with coefficients converted back to raw-scale-equivalent units for reporting; this puts each coefficient on the same log-odds-per-SD scale, so a shared prior places comparable mass across all four $\beta$s rather than being more restrictive for some than others.

Topic-specific coefficients are assigned independent Gaussian priors, centred at zero:

\[
\beta_\conv[t],\ \beta_\pers[t],\ \beta_\expo[t],\ \beta_{\conv \expo}[t] \sim \mathcal{N}(0,1), \quad \forall t\in\{1,\dots,K\}.
\]

On this scale, $\mathcal{N}(0,1)$ places about 95\% of its mass on effects between $-2$ and $2$ log-odds per one-SD change in the predictor, regularizing topic-level estimates toward zero while still admitting the larger effects reported in Section~\ref{subsec:primary-drivers} (several exceeding $|\beta|=3$).

User and debate intercepts are modelled as Gaussian random effects:

\[
\alpha_i \sim \mathcal{N}(0,\sigma_{\text{user}}), \quad
\delta_j \sim \mathcal{N}(0,\sigma_{\text{debate}}),
\]

with half-normal hyperpriors:

\[
\sigma_{\text{user}}, \sigma_{\text{debate}} \sim \mathcal{N}^+(0,1).
\]

A non-centred parameterisation is used for all hierarchical components to improve sampling efficiency \cite{betancourt2015hamiltonian}. Posterior inference is performed using Hamiltonian Monte Carlo with the No-U-Turn Sampler (NUTS)~\cite{hoffman2014nuts}, running four chains with 1,000 warm-up and 1,000 sampling iterations each. Pointwise log-likelihoods are computed to enable leave-one-out cross-validation (LOO-CV)~\cite{silva2024robust} for model comparison across nested specifications: the full model (Eq.~\ref{eq:model}); a no-interaction model (setting $\beta_{\conv \expo}[t]\equiv 0$); and three single-predictor models retaining only $\conv$, only $\expo$, or only $\pers$. All results reported below use this standardized-predictor specification.

\subsection{Summarizing the results}\label{subsec:met-param-class}
The model above yields 47 topic-specific posterior distributions, which can be difficult to interpret directly in relation to my research questions, i.e.,  \textit{Which is the stronger predictor of voting: conviction or persuasiveness? (\textbf{RQ1}) Does cross-cutting exposure change the effect of conviction? (\textbf{RQ2}).} To summarize these results, I classify the posterior evidence as follows.

\paragraph{RQ1: Which is the stronger predictor of voting: conviction or persuasiveness?} I first assess the \emph{strength} of the conviction and persuasiveness coefficients. A coefficient is considered \emph{strong} when the entire 90\% credible interval of its odds ratio $\mathrm{OR}=e^\beta$ excludes the null ($\mathrm{\beta}=0$), in either direction. Odds ratios are computed on raw-scale-equivalent coefficients, i.e., each standardized coefficient divided by the SD of its own predictor, so that strength refers to a one-unit increase in that predictor's own units. I then assess the \emph{direction} of dominance via the posterior distribution of $\Delta_t = \beta_\conv[t]-\beta_\pers[t]$. Since conviction and persuasiveness live on non-comparable raw scales, $\Delta_t$ is computed on the standardized coefficients, i.e., the effect of a one-SD change in each predictor, which is the scale on which the two are comparable. Direction is considered clear when the 90\% CI of $\Delta_t$ excludes zero: conviction dominates if $\Delta_{t,5\%}>0$, persuasiveness dominates if $\Delta_{t,95\%}<0$. Combining strength and direction yields the six summary labels in Table~\ref{tab:classification-primary}: \emph{conviction-driven} or \emph{persuasiveness-driven} when both coefficients are strong, and $\Delta_t$ clearly favours one side; \emph{jointly-driven} when both are strong but neither dominates; \emph{conviction-only} or \emph{persuasiveness-only} when just one coefficient is strong; and \emph{indeterminate} when neither is strong.

\paragraph{RQ2: Does cross-cutting exposure change the effect of conviction?} Independently of the above, I assess whether cross-cutting exposure moderates conviction via the posterior distribution of $\beta_{\conv \expo}[t]$, the topic-specific interaction coefficient. A topic is classified as \emph{exposure-moderated} when the 90\% CI of $\beta_{\conv \expo}[t]$ lies entirely below zero, consistent with the cross-cutting exposure account that a more ideologically diverse prior audience weakens reliance on conviction~\cite{mutz2006hearing}; as \emph{exposure-entrenched} when the 90\% CI lies entirely above zero, consistent with reactance or entrenchment accounts~\cite{bail2018exposure}; and as \emph{exposure-neutral} when the CI straddles zero, indicating conviction's effect is robust to the diversity of the prior audience for that topic (Table~\ref{tab:classification-moderation}). Each topic therefore receives one label from each classification. Both schemes are intended as a heuristic summary of posterior evidence rather than a formal decision rule.

\begin{table}[ht]
\centering
\caption{RQ1 summary labels: conviction ($\conv$) vs. persuasiveness ($\pers$). \emph{Strong}: full 90\% CI of the raw-scale-equivalent odds ratio excludes 1. $\Delta_t = \beta_\conv[t] - \beta_\pers[t]$, computed on standardized coefficients.}
\label{tab:classification-primary}
\small
\begin{tabular}{lccc}
\toprule
\textbf{RQ1 summary labels} & $\mathrm{OR}_\conv$ strong & $\mathrm{OR}_\pers$ strong & \textbf{90\% CI of $\Delta_t$} \\
\midrule
Conviction-driven  & \checkmark & \checkmark & $\Delta_{t,5\%} > 0$ \\
persuasiveness-driven  & \checkmark & \checkmark & $\Delta_{t,95\%} < 0$ \\
Jointly-driven     & \checkmark & \checkmark & straddles zero \\
Conviction-only    & \checkmark & $\times$   & --- \\
persuasiveness-only    & $\times$   & \checkmark & --- \\
Indeterminate      & $\times$   & $\times$   & --- \\
\bottomrule
\end{tabular}
\end{table}

\begin{table}[ht]
\centering
\caption{RQ2 summary labels: does cross-cutting exposure moderate conviction? \emph{Credible}: 90\% CI of $\beta_{\conv \expo}[t]$ excludes zero.}
\label{tab:classification-moderation}
\small
\begin{tabular}{lcl}
\toprule
\textbf{RQ2 summary labels} & \textbf{90\% CI of $\beta_{\conv \expo}[t]$} & \textbf{Theoretical reading} \\
\midrule
Exposure-moderated  & credibly $< 0$ & Mutz-consistent: diversity weakens conviction \\
Exposure-entrenched & credibly $> 0$ & Reactance/entrenchment-consistent \\
Exposure-neutral    & straddles zero & Conviction robust to audience diversity \\
\bottomrule
\end{tabular}
\end{table}
\clearpage
\section{Results}\label{sec:results}
\subsection{Model fit}\label{subsec:model-fit-interpr}
The full model (Eq.~\ref{eq:model}), incorporating conviction ($\conv$), persuasiveness ($\pers$), cross-cutting exposure ($\expo$), and their interaction ($\conv \times \expo$), together with user and debate intercepts ($\alpha$ and $\delta$), is compared against three nested specifications via leave-one-out cross-validation (LOO-CV; see Appendix~\ref{app:model-comparison} and Table~\ref{tab:model-comparison}). The three-predictor model ($\conv+\expo+\pers$) is the best-fitting specification, with $\mathrm{ELPD}=-25{,}700$ and $89\%$ of the model weight. The interaction model fits slightly worse ($-25{,}700$, $\Delta\mathrm{ELPD}\approx-20$, $\Delta\mathrm{SE}=6.7$) and takes only 1\% of the weight. The $\conv\times \expo$ interaction therefore does not improve out-of-sample predictive fit beyond the three main effects, even though its coefficient is credibly non-zero on a handful of individual topics, as I report in Section~\ref{subsec:moderation-results}. Thus, the evidence for topic-specific moderation should not be interpreted as evidence that the interaction materially improves prediction at the model level.

\begin{figure}
    \centering
    \begin{subfigure}[b]{\linewidth}
        \centering
        \includegraphics[width=\linewidth]{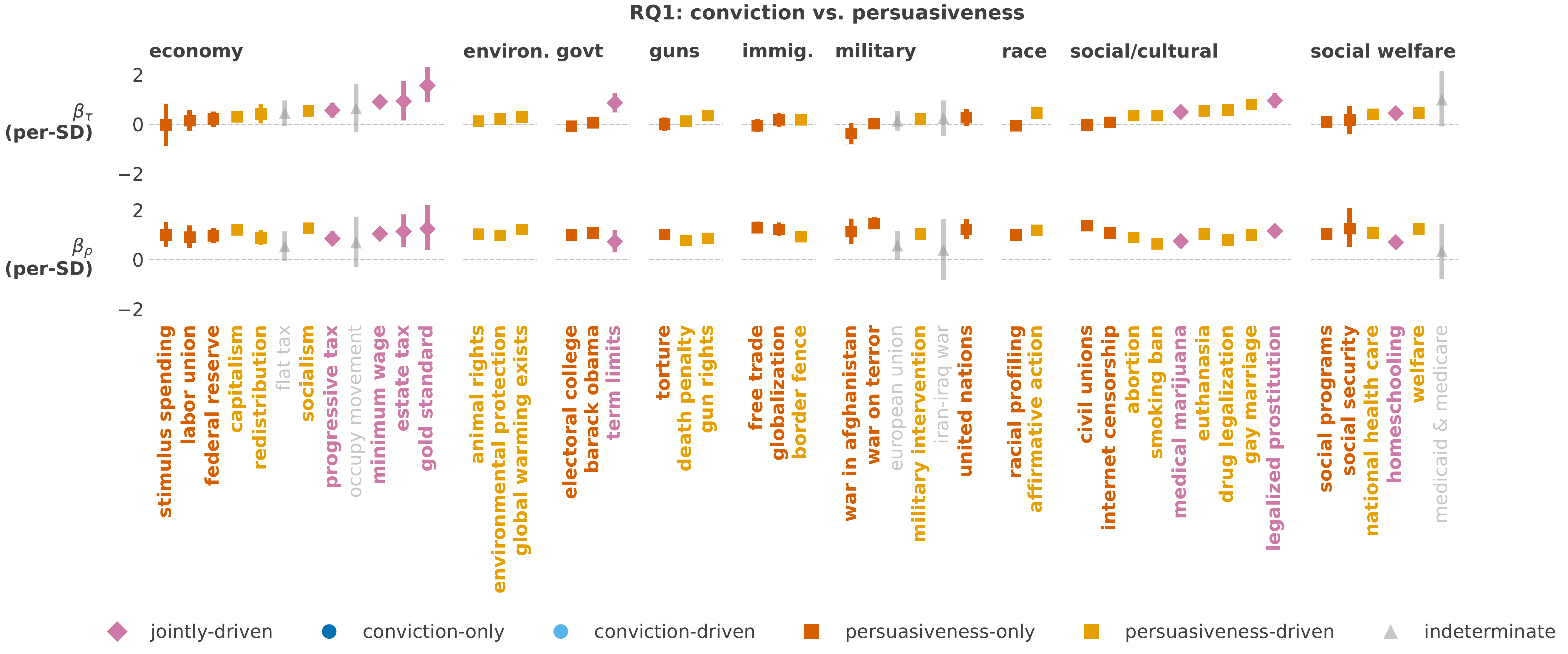}
        \caption{RQ1: conviction ($\beta_\conv$) vs. persuasiveness ($\beta_\pers$).}
        \label{fig:params-forest-rq1}
    \end{subfigure}

    \bigskip

    \begin{subfigure}[b]{\linewidth}
        \centering
        \includegraphics[width=\linewidth]{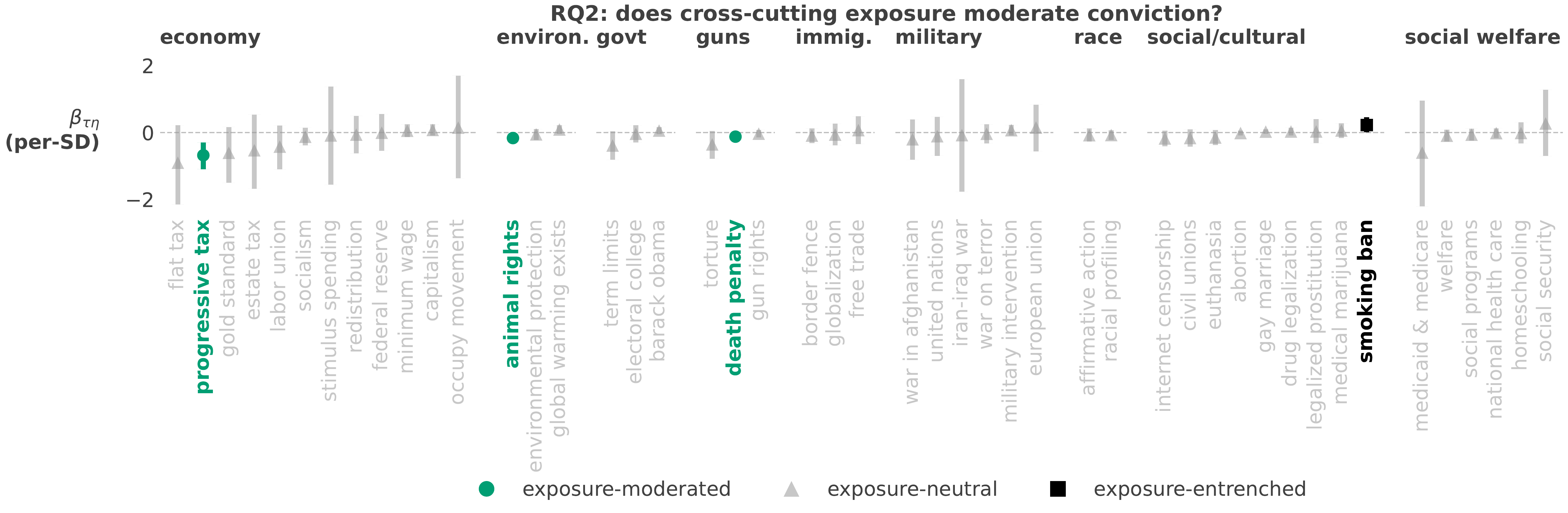}
        \caption{RQ2: cross-cutting exposure moderation of conviction ($\beta_{\conv \expo}$).}
        \label{fig:params-forest-rq2}
    \end{subfigure}

    \caption{\textbf{Topic-specific parameter estimates.} Posterior means and 90\% credible
    intervals across all 47 topics, grouped by issue and sorted by point estimate within
    each issue. \textbf{(a)} $\beta_\conv$ (conviction) and $\beta_\pers$ (persuasiveness),
    coloured by RQ1 summary labels. \textbf{(b)} $\beta_{\conv \expo}$
    (conviction $\times$ cross-cutting exposure), coloured by RQ2 summary labels.
    Bold labels indicate topics whose full 90\% credible interval of the raw-scale-equivalent
    odds ratio excludes the null (threshold $=1.0$; see
    Section~\ref{subsec:met-param-class}) for the respective coefficient.}
    \label{fig:params-forest}
\end{figure}

Before turning to the 47 topic-specific estimates, it is useful to summarize their overall distribution. Across topics, conviction coefficients average $\hat\beta_\tau=1.95$ (median $1.61$, range $-2.04$ to $8.66$), with 26 of 47 topics showing a 90\% credible interval that excludes zero; no topic shows a credibly negative conviction effect. Persuasiveness coefficients average $\hat\beta_\pers=1.61$ (median $1.66$, range $0.54$ to $2.39$) and are strikingly uniform in direction: every one of the 47 point estimates is positive, and 42 of 47 credibly so. The conviction$\times$exposure interaction is more diffuse: its median is $-2.38$, but its mean ($-5.22$) is pulled well below the median by a small number of very large, wide-interval estimates on sparsely covered topics (Section~\ref{subsec:moderation-results}); only 4 of 47 topics show a credible interaction in either direction. This overall picture, i.e., persuasiveness as a reliably positive, near-universal predictor, conviction as a positive but more variable secondary predictor, and the interaction term as largely noise except on a handful of topics, is developed at the topic level below. Posterior means and 90\% credible intervals for all topic-specific coefficients ($\beta_\conv$, $\beta_\pers$, $\beta_{\conv \expo}$) are reported in Table~\ref{tab:results-estimation} and visualised in Figure~\ref{fig:params-forest}.

\scriptsize
\begin{longtable}{llllll}
\caption{\textbf{Posterior summaries for topic-specific coefficients (raw-scale-equivalent units).} For each topic, $\hat\beta_\conv$ and $\hat\beta_\pers$ report the conviction and persuasiveness coefficients (with 90\% credible intervals) used to assign the RQ1 summary label in the adjacent column, following the classification in Table~\ref{tab:classification-primary}; $\hat\beta_{\conv \expo}$ reports the conviction $\times$ cross-cutting-exposure interaction coefficient used to assign the RQ2 summary label, following the classification in Table~\ref{tab:classification-moderation}. Bold indicates a coefficient whose 90\% CI excludes zero, i.e.\ \emph{strong} under the criteria in Section~\ref{subsec:met-param-class}. Topics sorted alphabetically.
\label{tab:results-estimation}}\\
\toprule
topic & $\hat\beta_\conv$ [90\% CI] & $\hat\beta_\pers$ [90\% CI] & RQ1 summary labels & $\hat\beta_{\conv \expo}$ [90\% CI] & RQ2 summary labels \\
\midrule
\endfirsthead
\midrule
topic & $\hat\beta_\conv$ [90\% CI] & $\hat\beta_\pers$ [90\% CI] & RQ1 summary labels & $\hat\beta_{\conv \expo}$ [90\% CI] & RQ2 summary labels \\
\midrule
\endhead
\bottomrule
\endfoot
abortion & \textbf{1.96} [1.71, 2.21] & \textbf{1.45} [1.35, 1.56] & persuasiveness-driven & -0.40 [-3.58, 2.86] & exposure-neutral \\
affirmative action & \textbf{2.49} [1.57, 3.45] & \textbf{1.94} [1.64, 2.25] & persuasiveness-driven & -4.33 [-13.19, 4.75] & exposure-neutral \\
animal rights & \textbf{0.68} [0.34, 1.03] & \textbf{1.67} [1.55, 1.80] & persuasiveness-driven & \textbf{-8.64} [-14.62, -2.89] & exposure-moderated \\
barack obama & 0.37 [-0.20, 0.93] & \textbf{1.76} [1.60, 1.92] & persuasiveness-only  & 2.19 [-3.08, 7.33] & exposure-neutral \\
border fence & \textbf{1.05} [0.28, 1.79] & \textbf{1.53} [1.23, 1.84] & persuasiveness-driven  & -3.76 [-13.94, 5.94] & exposure-neutral \\
capitalism & \textbf{1.71} [1.15, 2.28] & \textbf{1.97} [1.78, 2.17] & persuasiveness-driven & 3.21 [-5.09, 11.29] & exposure-neutral \\
civil unions & -0.16 [-1.12, 0.81] & \textbf{2.26} [1.96, 2.57] & persuasiveness-only & -8.59 [-20.69, 3.70] & exposure-neutral \\
death penalty & \textbf{0.65} [0.29, 1.02] & \textbf{1.27} [1.12, 1.42] & persuasiveness-driven & \textbf{-5.68} [-10.04, -1.24] & exposure-moderated \\
drug legalization & \textbf{3.23} [2.74, 3.72] & \textbf{1.31} [1.14, 1.48] & persuasiveness-driven & 1.56 [-3.63, 6.81] & exposure-neutral \\
electoral college & -0.43 [-1.41, 0.55] & \textbf{1.62} [1.35, 1.90] & persuasiveness-only & -1.97 [-14.40, 9.95] & exposure-neutral \\
environmental protection & \textbf{1.19} [0.54, 1.85] & \textbf{1.61} [1.38, 1.84] & persuasiveness-driven  & -1.66 [-9.19, 5.93] & exposure-neutral \\
estate tax & \textbf{5.16} [0.86, 9.67] & \textbf{1.86} [0.85, 2.99] & jointly-driven & -19.80 [-74.77, 32.75] & exposure-neutral \\
european union & 0.77 [-1.43, 3.01] & 0.92 [-0.01, 1.90] & indeterminate & 6.29 [-27.00, 38.84] & exposure-neutral \\
euthanasia & \textbf{3.01} [2.25, 3.78] & \textbf{1.70} [1.45, 1.95] & persuasiveness-driven & -6.60 [-17.17, 3.92] & exposure-neutral \\
federal reserve & 1.13 [-0.54, 2.85] & \textbf{1.57} [1.07, 2.11] & persuasiveness-only & 0.18 [-25.48, 26.45] & exposure-neutral \\
flat tax & 2.53 [-0.35, 5.33] & 0.87 [-0.08, 1.86] & indeterminate & -43.83 [-100.96, 8.36] & exposure-neutral \\
free trade & -0.29 [-1.77, 1.23] & \textbf{2.12} [1.74, 2.52] & persuasiveness-only & 4.42 [-14.88, 23.06] & exposure-neutral \\
gay marriage & \textbf{4.43} [4.13, 4.74] & \textbf{1.62} [1.53, 1.72] & persuasiveness-driven & 0.62 [-2.60, 3.74] & exposure-neutral \\
global warming exists & \textbf{1.61} [1.06, 2.16] & \textbf{1.99} [1.82, 2.18] & persuasiveness-driven & 4.18 [-1.73, 10.13] & exposure-neutral \\
globalization & 1.01 [-0.54, 2.61] & \textbf{2.00} [1.57, 2.46] & persuasiveness-only & -3.36 [-18.37, 11.82] & exposure-neutral \\
gold standard & \textbf{8.66} [4.88, 12.76] & \textbf{2.04} [0.64, 3.60] & jointly-driven & -23.74 [-63.78, 10.51] & exposure-neutral \\
gun rights & \textbf{1.93} [1.46, 2.38] & \textbf{1.40} [1.23, 1.57] & persuasiveness-driven & -1.97 [-7.54, 3.37] & exposure-neutral \\
homeschooling & \textbf{2.47} [1.40, 3.54] & \textbf{1.15} [0.84, 1.46] & jointly-driven  & -1.17 [-16.14, 13.59] & exposure-neutral \\
internet censorship & 0.39 [-0.43, 1.24] & \textbf{1.75} [1.53, 1.98] & persuasiveness-only & -8.04 [-18.90, 2.41] & exposure-neutral \\
iran-iraq war & 1.35 [-2.63, 5.29] & 0.64 [-1.35, 2.70] & indeterminate & 0.41 [-76.92, 77.18] & exposure-neutral \\
labor union & 0.85 [-1.41, 3.18] & \textbf{1.48} [0.77, 2.26] & persuasiveness-only  & -24.99 [-57.50, 6.98] & exposure-neutral \\
legalized prostitution & \textbf{5.29} [3.69, 6.93] & \textbf{1.89} [1.52, 2.28] & jointly-driven & 1.18 [-15.34, 18.28] & exposure-neutral \\
medicaid \& medicare & 5.51 [-0.46, 11.89] & 0.54 [-1.27, 2.36] & indeterminate & -25.65 [-98.62, 47.79] & exposure-neutral \\
medical marijuana & \textbf{2.76} [1.45, 4.07] & \textbf{1.22} [0.78, 1.64] & jointly-driven & 4.04 [-6.54, 14.15] & exposure-neutral \\
military intervention & \textbf{1.16} [0.56, 1.76] & \textbf{1.70} [1.55, 1.85] & persuasiveness-driven & 2.96 [-4.39, 10.41] & exposure-neutral \\
minimum wage & \textbf{5.02} [3.97, 6.08] & \textbf{1.71} [1.36, 2.06] & jointly-driven & 1.98 [-7.37, 10.84] & exposure-neutral \\
national health care & \textbf{2.20} [1.48, 2.92] & \textbf{1.77} [1.54, 2.00] & persuasiveness-driven & -0.75 [-7.67, 5.93] & exposure-neutral \\
occupy movement & 3.54 [-1.77, 9.05] & 1.13 [-0.50, 2.84] & indeterminate & 12.70 [-65.99, 93.09] & exposure-neutral \\
progressive tax & \textbf{3.13} [1.51, 4.79] & \textbf{1.39} [0.94, 1.88] & jointly-driven & \textbf{-32.57} [-52.57, -14.92] & exposure-moderated \\
racial profiling & -0.33 [-0.99, 0.28] & \textbf{1.63} [1.42, 1.83] & persuasiveness-only & -3.58 [-10.29, 2.98] & exposure-neutral \\
redistribution & \textbf{2.29} [0.21, 4.46] & \textbf{1.45} [0.97, 1.94] & persuasiveness-driven  & -4.91 [-33.67, 23.69] & exposure-neutral \\
smoking ban & \textbf{1.98} [1.07, 2.90] & \textbf{1.06} [0.78, 1.33] & persuasiveness-driven & \textbf{11.80} [0.78, 23.63] & exposure-entrenched \\
social programs & 0.57 [-0.20, 1.34] & \textbf{1.71} [1.45, 1.97] & persuasiveness-only & -2.38 [-10.49, 5.59] & exposure-neutral \\
social security & 0.93 [-2.23, 4.10] & \textbf{2.06} [0.84, 3.42] & persuasiveness-only & 12.26 [-37.04, 62.44] & exposure-neutral \\
socialism & \textbf{2.98} [2.00, 3.96] & \textbf{2.06} [1.74, 2.41] & persuasiveness-driven & -6.72 [-18.67, 5.16] & exposure-neutral \\
stimulus spending & -0.12 [-4.88, 4.61] & \textbf{1.63} [0.83, 2.50] & persuasiveness-only & -13.57 [-90.38, 62.71] & exposure-neutral \\
term limits & \textbf{4.79} [2.62, 6.96] & \textbf{1.19} [0.48, 1.95] & jointly-driven & -16.86 [-36.86, 2.90] & exposure-neutral \\
torture & 0.05 [-1.43, 1.55] & \textbf{1.66} [1.33, 2.00] & persuasiveness-only & -15.82 [-35.40, 3.43] & exposure-neutral \\
united nations & 1.47 [-0.42, 3.35] & \textbf{2.00} [1.36, 2.67] & persuasiveness-only & -5.99 [-32.26, 19.49] & exposure-neutral \\
war in afghanistan & -2.04 [-4.48, 0.30] & \textbf{1.86} [1.05, 2.71] & persuasiveness-only & -11.08 [-38.60, 16.33] & exposure-neutral \\
war on terror & 0.16 [-0.80, 1.07] & \textbf{2.39} [2.00, 2.80] & persuasiveness-only & -1.86 [-14.42, 11.39] & exposure-neutral \\
welfare & \textbf{2.51} [1.63, 3.38] & \textbf{2.02} [1.74, 2.33] & persuasiveness-driven & -4.83 [-13.36, 3.60] & exposure-neutral \\
\end{longtable}
\normalsize

\subsection{RQ1: conviction versus persuasiveness}\label{subsec:primary-drivers}
Under the classification scheme in Table~\ref{tab:classification-primary}, of the 47 classifiable topics, 18 are \emph{persuasiveness-driven}, 16 are \emph{persuasiveness-only}, 8 are \emph{jointly-driven}, and 5 are \emph{indeterminate} (see Fig.~\ref{fig:params-forest-rq1}). No topic is classified as \emph{conviction-driven} or \emph{conviction-only}. Persuasiveness has a strong, credible effect on 42 of 47 topics (18 persuasiveness-driven, 16 persuasiveness-only, and 8 jointly-driven), and conviction on 26 (18 persuasiveness-driven and 8 jointly-driven), i.e., conviction is never strong on a topic where persuasiveness is not also strong. On no topic is conviction credibly larger than persuasiveness, and on 18 topics persuasiveness is credibly larger. The following subsections describe each summary label in turn.

\subsubsection{Persuasiveness-driven and persuasiveness-only topics}
Eighteen topics have both coefficients credibly non-null, with persuasiveness the larger of the two (\emph{persuasiveness-driven})~\footnote{persuasiveness-driven: \textit{abortion}, \textit{affirmative action}, \textit{animal rights}, \textit{border fence}, \textit{capitalism}, \textit{death penalty}, \textit{drug legalization}, \textit{environmental protection}, \textit{euthanasia}, \textit{gay marriage}, \textit{global warming exists}, \textit{gun rights}, \textit{military intervention}, \textit{national health care}, \textit{redistribution}, \textit{smoking ban}, \textit{socialism}, and \textit{welfare}}. A further 16 topics are \emph{persuasiveness-only}, i.e., they show a strong persuasiveness effect and no comparably strong conviction effect~\footnote{persuasiveness-only:\textit{barack obama}, \textit{civil unions}, \textit{electoral college}, \textit{federal reserve}, \textit{free trade}, \textit{globalization}, \textit{internet censorship}, \textit{labor union}, \textit{racial profiling}, \textit{social programs}, \textit{social security}, \textit{stimulus spending}, \textit{torture}, \textit{united nations}, \textit{war in afghanistan}, and \textit{war on terror}}. These 34 topics (72\% of the dataset) span all nine ANES issue groupings (Table~\ref{tab:topic-issues}). Several of these topics have a large $\hat\beta_\conv$ in raw-scale-equivalent units and are still classified as persuasiveness-driven, e.g., \textit{drug legalization} ($\hat\beta_\conv=3.23$ against $\hat\beta_\pers=1.31$) and \textit{gay marriage}. This follows from how dominance is assessed: $\Delta_t$ is computed on the standardized scale, on which the two coefficients are comparable (Section~\ref{subsec:met-param-class}), and on that scale it credibly favours persuasiveness.

Substantively, persuasiveness extends well beyond the geopolitical and institutional topics (barack obama, electoral college, federal reserve, war on terror), into topics with an obvious personal stake. These include canonical body-autonomy and lifestyle topics (abortion, drug legalization, euthanasia, gay marriage) that the moral conviction literature expects to be resistant to influence~\cite{skitka2010psychology,skitka2021psychology}.

This raises a question about the mechanism behind $\pers_{ij}$ itself. A voter's persuasiveness judgement may be partly downstream of their own prior stance, and not an independent read of argument quality; if so, part of persuasiveness's apparent dominance would reflect conviction operating through the persuasiveness channel. I test this directly among genuine third-party voters who both report a prior belief on the debate's topic ($\conv_{ij}\neq0$) and give a substantive persuasiveness judgement ($\pers_{ij}\neq0$; $N=11{,}032$, excluding debaters and ties/no-preference judgements). Conviction and persuasiveness are positively correlated ($r=0.14$, $p\approx7\times10^{-49}$), and these voters credit the debater on their own prior-belief-aligned side 56.3\% of the time, against a 50\% chance baseline. The confound is modest in size: $\conv$ and $\pers$ share only about 2\% of variance, so persuasiveness's overall dominance stays valid. Some of what the model attributes to independent argument quality is more plausibly prior belief expressing itself through the persuasiveness channel, particularly on the body-autonomy and lifestyle topics flagged above.

\subsubsection{Jointly-driven topics}
Eight topics have both conviction and persuasiveness credibly strong, with no clear dominance between them: \textit{estate tax}, \textit{gold standard}, \textit{homeschooling}, \textit{legalized prostitution}, \textit{medical marijuana}, \textit{minimum wage}, \textit{progressive tax}, and \textit{term limits}. This is the summary label where conviction contributes similarly to persuasiveness, and several of its topics (legalized prostitution, medical marijuana, homeschooling) fit the personal-freedom and lifestyle profile that the moral conviction literature predicts~\cite{skitka2010psychology,skitka2021psychology}. Persuasiveness is credibly non-null on all eight, however, so conviction here adds to persuasiveness and does not displace it.

\subsubsection{Indeterminate topics}\label{subsubsec:indet-topics}
Five topics show neither a strong conviction effect nor a strong persuasiveness effect: \textit{european union}, \textit{flat tax}, \textit{iran-iraq war}, \textit{medicaid \& medicare}, and \textit{occupy movement}. All five have sparse coverage: they have 5 (medicaid \& medicare), 7 (occupy movement), 14 (iran-iraq war), 28 (flat tax), and 44 (european union) debates in the modelled dataset, i.e., all five rank among the eight lowest-coverage topics in the dataset, against a median of 462 debates for persuasiveness-driven topics. Appendix~\ref{app:user-demographics}, Figure~\ref{fig:users-prior-beliefs}, corroborates this from the users' side: few users report a prior stance on several of these topics. Four of the five also fall in \textit{economy, budget, and taxes} or \textit{military and foreign policy}, the two issue areas with the largest topic counts in Table~\ref{tab:topic-issues}, i.e., broad areas where debate volume is spread thin across many constituent topics. The \emph{indeterminate} label therefore reflects a lack of statistical power.

\subsection{RQ2: Does cross-cutting exposure moderate conviction?}\label{subsec:moderation-results}
Turning to the second research question of this paper, Table~\ref{tab:results-estimation} classifies topics by the credibility and direction of $\beta_{\conv \expo}[t]$. Cross-cutting exposure credibly moderates conviction on 4 of 47 topics (9\%): 3 \emph{exposure-moderated} and 1 \emph{exposure-entrenched}, with the remaining 43 (91\%) \emph{exposure-neutral} (see Fig.~\ref{fig:params-forest-rq2}). %At a per-topic 90\% credible interval across 47 independently estimated topics, four excluded-zero intervals are close to what chance alone would produce, so I treat these as candidates for moderation rather than confirmed effects, and weigh each against its debate coverage below.

Three topics show a nominally credible negative $\beta_{\conv \expo}$, consistent with Mutz's moderation account: \textit{animal rights} ($\hat\beta_{\conv \expo}=-8.64$, 90\% CI $[-14.62,-2.89]$), \textit{death penalty} ($\hat\beta_{\conv \expo}=-5.68$, CI $[-10.04,-1.24]$), and \textit{progressive tax} ($\hat\beta_{\conv \expo}=-32.57$, CI $[-52.57,-14.92]$). Conviction is independently strong on all three (Section~\ref{subsec:primary-drivers}): animal rights and death penalty are \emph{persuasiveness-driven} ($\hat\beta_\conv=0.68$ and $0.65$), and progressive tax is \emph{jointly-driven} ($\hat\beta_\conv=3.13$, CI $[1.51,4.79]$). Progressive tax has both the most extreme point estimate and the widest CI of the three, and has a 74--142$\times$ sparser coverage (9 debates, versus 1{,}278 for animal rights and 669 for death penalty). The topic with the strongest apparent moderation effect is thus also the one most likely, by the coverage heuristic, to carry an inflated estimate.

One topic, \textit{smoking ban}, shows a nominally credible positive $\beta_{\conv \expo}$ ($\hat\beta_{\conv \expo}=11.80$, CI $[0.78,23.63]$), consistent with entrenchment/reactance accounts~\cite{bail2018exposure}. It is \emph{persuasiveness-driven} ($\hat\beta_\conv=1.98$, CI $[1.07,2.90]$), so conviction is independently established, though secondary to persuasiveness. With its CI lower bound close to zero and only one topic of 47 showing this pattern, I report it as a suggestive, low-confidence exception and cautiously evaluate its evidence for reactance accounts.

The remaining 43 topics (91\%) show no credible relationship between prior-audience diversity and conviction's pull on the vote. CI width for $\beta_{\conv \expo}$ is strongly, negatively related to debate coverage across all 47 topics (Spearman $r=-0.95$), so most of this null reflects sparse coverage rather than a genuinely well-estimated absence of effect. Only the most densely covered topics, i.e., gay marriage, abortion, gun rights, drug legalization, barack obama, global warming exists, capitalism, military intervention, racial profiling, environmental protection, and national health care (all $>450$ debates, CI width $<17$ raw-scale-equivalent units) represent a well-powered null; the rest are underpowered and do not constitute evidence against moderation.

Cross-cutting exposure theory, as operationalised by $\expo_{ij}$ here, therefore finds only narrow, topic-specific support in this dataset, concentrated on topics whose coverage makes the estimates hard to trust. I return to this in Section~\ref{sec:limitations}.
\subsection{Heterogeneity across issues}\label{subsec:macro-heterogeneity}
Table~\ref{tab:macro-primary} and Table~\ref{tab:macro-moderation} cross-tabulate the two classification schemes against the nine ANES issue groupings from Table~\ref{tab:topic-issues}.

\begin{table}[htbp]
\centering
\small
\caption{\textbf{RQ1 summary labels by issue area.}}
\label{tab:macro-primary}
\begin{tabular}{lcccc}
\toprule
Issue & persuasiveness-driven & Jointly-driven & persuasiveness-only & Indeterminate \\
\midrule
Economy, budget, and taxes                & 3 & 4 & 3 & 2 \\
Environment                                & 3 & 0 & 0 & 0 \\
Government and politics                    & 0 & 1 & 2 & 0 \\
Guns and crime                             & 2 & 0 & 1 & 0 \\
Immigration, trade, and globalization      & 1 & 0 & 2 & 0 \\
Military and foreign policy                & 1 & 0 & 3 & 2 \\
Race                                       & 1 & 0 & 1 & 0 \\
Social and cultural issues                 & 5 & 2 & 2 & 0 \\
Social welfare, healthcare, and education  & 2 & 1 & 2 & 1 \\
\bottomrule
\end{tabular}
\end{table}

\begin{table}[htbp]
\centering
\small
\caption{\textbf{RQ2 summary labels by issue area.}}
\label{tab:macro-moderation}
\begin{tabular}{lccc}
\toprule
Issue & Exposure-entrenched & Exposure-moderated & Exposure-neutral \\
\midrule
Economy, budget, and taxes                & 0 & 1 & 11 \\
Environment                                & 0 & 1 & 2 \\
Government and politics                    & 0 & 0 & 3 \\
Guns and crime                             & 0 & 1 & 2 \\
Immigration, trade, and globalization      & 0 & 0 & 3 \\
Military and foreign policy                & 0 & 0 & 6 \\
Race                                       & 0 & 0 & 2 \\
Social and cultural issues                 & 1 & 0 & 8 \\
Social welfare, healthcare, and education  & 0 & 0 & 6 \\
\bottomrule
\end{tabular}
\end{table}

For the \textbf{RQ1} labelling, \textit{environment} is the clearest outlier: all 3 of its topics (animal rights, environmental protection, global warming exists) are persuasiveness-driven, and it is the only issue area with no jointly-driven, persuasiveness-only, or indeterminate topics. \textit{Social and cultural issues} show the most balanced split between persuasiveness-driven (5 of 9: abortion, drug legalization, euthanasia, gay marriage, smoking ban) and jointly-driven (2 of 9: legalized prostitution, medical marijuana). This is where conviction appears most, consistent with the grouping's mix of body-autonomy and lifestyle topics, although persuasiveness is never credibly smaller than conviction even there. \textit{Government and politics} is the only issue area with no persuasiveness-driven topics (1 jointly-driven, 2 persuasiveness-only), while \textit{immigration, trade, and globalization} and \textit{race} are almost entirely persuasiveness-driven or persuasiveness-only, with no jointly-driven or indeterminate topics. \textit{Economy, budget, and taxes} is the largest issue group (12 topics, after excluding the unclassified national retail sales tax) and the most internally heterogeneous: it contributes topics to all four classes, and holds the highest concentration of \textit{jointly-driven} topics (4 of 12: estate tax, gold standard, minimum wage, progressive tax), alongside 2 of the 5 indeterminate topics (flat tax, occupy movement). This is consistent with the sparse-coverage explanation discussed in Section~\ref{subsec:primary-drivers} for some of its constituent topics.

For the \textbf{RQ2} labelling, the pattern is much flatter, as expected given the lopsided overall split (43 of 47 exposure-neutral). The 3 exposure-moderated topics (animal rights, death penalty, progressive tax) fall in three different issue areas (environment; guns and crime; economy, budget, and taxes), and the single exposure-entrenched topic (smoking ban) sits in \textit{social and cultural issues}. No issue area shows more than one nominally moderated topic. This dataset therefore does not support a claim that cross-cutting exposure operates differently by issue domain: where the moderation effect appears at all, it is topic-specific, weakly identified, and not issue-general.

\section{Discussion}\label{sec:conclusion}
\subsection{Persuasiveness, conviction, and the deliberative ideal}
Every choice we make is shaped by a complex interplay of forces: the values instilled by our upbringing, the social circles we inhabit, and the information environments we navigate. Online platforms have added a new layer to this complexity, creating spaces where personal beliefs and social signals are constantly entangled. \textit{Debate.org} was a particularly interesting example of such an ecosystem: users voted on controversial topics while seeing both the arguments made and the audience that had already voted, which makes it a rare setting in which conviction, persuasiveness, and cross-cutting exposure can be observed and separated at once.

Answering the two questions this paper set out to ask: persuasiveness is the stronger and more consistent predictor of vote choice, credible on 42 of 47 topics against 26 for conviction, which it never credibly outweighs, though on 8 topics, several tied to personal freedom and lifestyle, conviction's contribution is comparable in size. Cross-cutting exposure, by contrast, moderates conviction's pull on only 4 of 47 topics, a null that mostly reflects sparse coverage rather than a well-powered absence of effect. The central contribution of this work is an empirical decomposition of all three mechanisms within a single estimable framework, making it possible to see not just which predictor dominates on average, but how that balance shifts topic by topic.

Read against the deliberative democracy described by Chambers~\cite{chambers2025deliberation}, this is an interesting result, and it maps onto the dual-process distinction between systematic evaluation of argument content and heuristic reliance on cues~\cite{chaiken1980heuristic,petty1986elaboration}. Votes are better explained by the voter's assessment of the debate than by the stance they brought to it, and that ordering holds on 42 of 47 topics. Two qualifications keep the claim in proportion. The assessment and the vote are recorded on the same ballot, so their agreement reflects the coherence of a single judgement rather than two independent measurements; and the test in Section~\ref{subsec:primary-drivers} establishes that the assessment is not a restatement of prior belief (about 2\% shared variance) without establishing that it was formed by weighing the arguments. What the result does show, robustly, is that vote choice here is not a readout of the commitments voters arrive with. It does not, however, establish that voters changed their underlying beliefs as a result of evaluating the arguments.

The cross-cutting-exposure results do not support the theory’s prediction, but they may nevertheless be informative for understanding the relationship between exposure and voting. Conviction's pull on the vote is stable across audience compositions: on 43 of 47 topics, the ideological diversity of the audience a voter arrives in bears no credible relationship to how strongly that voter relies on prior belief. Conviction here behaves much as the moral conviction literature describes it, i.e., as robust to the composition of the surrounding crowd~\cite{skitka2021psychology,aramovich2012opposing}, and the four topics that do show moderation are best read as leads for future work given their debate coverage.

The implications for platform design are correspondingly constructive. If votes largely reproduced prior beliefs, the content of a debate would matter little to its outcome, and design interventions would have little to work with. One reading tempers this: a voter's sense of ideological closeness to a debater could itself shape how persuasive that debater seems, so prior belief may still be doing much of the work even where persuasiveness measures larger, a reason for some caution in reading the causal weight of debate content. These results point the other way, and the ordering holds across nearly every topic, including the morally charged ones, which suggests it is a property of the setting rather than of particular issues. The audience-composition results are weaker. A model containing only $\expo$ fits worst of the five specifications compared in Appendix~\ref{app:model-comparison}, and at the topic level the intervals are widest wherever debate coverage is thinnest. On most topics there is not enough data to tell a null apart from a moderate effect, so I would not read this as showing that audience composition does not matter. It does suggest that any risk here comes from voters seeing earlier votes before casting their own, not from the ballot itself.

\subsection{Moral conviction and personal-autonomy topics}
The moral conviction literature predicts that attitudes on issues of personal autonomy and lifestyle are experienced as objective facts and resist social influence~\cite{skitka2010psychology,skitka2021psychology,aramovich2012opposing,conover2018republicans}. These results both support that prediction and qualify it in an informative way. Supporting it, conviction is credible on several morality-policy topics~\cite{mooney2001clash}: legalized prostitution, medical marijuana, and homeschooling are among the eight jointly-driven topics; legalized prostitution, gay marriage, and drug legalization carry some of the largest $\hat\beta_\conv$ estimates among topics where conviction is credible. Qualifying it, persuasiveness is credibly at least as large as conviction even on these topics, and the largest estimates overall fall on tax and monetary-policy topics, which the moral conviction literature does not predict. Prior belief predicts votes on abortion, drug legalization, euthanasia, and gay marriage, and the assessment of who argued better predicts them more.

This is a finding worth pursuing, and it admits three readings that this design can begin to separate. First, the pattern may be specific to the task: debate.org voters are third-party observers judging a contest between two debaters, and not respondents stating their own position, so the setting may invite argument evaluation in a way a survey item does not. Second, moral conviction is a property of \emph{how} an individual holds a belief, while the platform records only its direction; a user who reports opposing drug legalization may or may not hold that position with moral conviction in Skitka's sense, so topic-level estimates likely understate the effect for the subset of users who do. Third, the small correlation reported in Section~\ref{subsec:primary-drivers} suggests part of persuasiveness's advantage on these topics operates through the persuasiveness judgement itself. A direct measure of moral conviction at the user-topic level would adjudicate between them, and collecting one is a natural next step.

\subsection{Scope and future directions}\label{sec:limitations}
Three features of the design bound the scope of these conclusions and suggest where the framework can be extended.

The first concerns coverage. Debate volume is uneven across topics, and this shapes how confidently individual topic-level coefficients can be read: the five indeterminate topics are among the eight lowest-coverage topics in the dataset, and the credible-interval width of $\beta_{\conv \expo}$ tracks coverage closely (Spearman $r=-0.95$). Topic coefficients are estimated under independent priors, so sparse topics receive no shrinkage. Hierarchical priors with partial pooling could stabilise estimates for sparse topics, and would let the framework scale to platforms with a longer tail of topics.

The second concerns measurement. Debate topics are inferred from text, and the resulting vectors are one-dimensional representations of richer content; richer, continuous topic representations would sharpen the conviction term. Likewise, $\pers_{ij}$ records a perceived judgement, so pairing it with text-based measures of argument quality, of the kind developed for related platforms~\cite{habernal2016convincing,tan2016winning}, would let central-route and peripheral-route persuasiveness be distinguished directly. A further extension would compare this approach against a fully latent ideal-point specification, in which user and debate positions are estimated jointly from voting behaviour. I do not pursue this here: with 15k users, 78k debates, and $K=48$ topics, a multidimensional latent-position model would add far more parameters than the present specification, likely straining identifiability given how sparse coverage already is for many topics (Section~\ref{subsubsec:indet-topics}). A lower-dimensional version would be more tractable, but would collapse the topic-by-topic heterogeneity that is this paper's central finding. On the exposure side, $\expo_{ij}$ captures diffuse heterogeneity well and cannot separate it from an evenly split two-camp audience (Appendix~\ref{app:diversity-simulation}), which matters most for the reactance reading~\cite{bail2018exposure}.

The third concerns interpretation. The estimated effects are conditional associations: users may select into debates whose audience already has a given composition, a version of the selection-versus-influence problem long noted in the political-network literature~\cite{manski1993}, and the placebo comparison in Appendix~\ref{app:diversity-placebo} is a partial check on it. Relatedly, the platform records each voter's stance both before and after the debate, which opens a direct extension of this framework to belief change itself: rather than asking which side a voter ends on, one can ask whether the debate moved them, and whether persuasiveness predicts movement. That is the sharpest available test of the deliberative claim this paper motivates.

Finally, the \textit{debate.org} user base skews young, male, white, and politically engaged (Appendix~\ref{app:user-demographics}, Figure~\ref{fig:demographics}). The framework itself transfers to any platform recording stances, argument judgements, and vote order; replication across such settings, particularly ones with more demographically balanced participation, would establish how far these specific estimates generalise.
\clearpage

%TC:ignore
\section*{Declarations}
\paragraph{Competing Interests}
None.

\paragraph{Funding}
EC acknowledges the support of the AccelNet-MultiNet program, a project of the National Science Foundation (Award \#1927425 and \#1927418).

\paragraph{Acknowledgments}
The author thanks Filippo Menczer and Alessandro Flammini for their hospitality and support during a visit to Indiana University, and is grateful to Erik-Jan van Kesteren, Javier Garcia-Bernardo, and Daniel Oberski for helpful discussions and feedback, as well as to Byunghwee Lee for pointing out the dataset.

\paragraph{Code Availability Statement}
The code associated with this research will be publicly available upon publication.

\paragraph{Ethical Statement}
Ethical approval for this study was obtained from the Faculty Ethics Review Board (FERB) of Utrecht University (Approval number 26-0191).

\paragraph{Declaration of generative AI and AI-assisted technologies in the writing process}
During the preparation of this work, the author used Claude Sonnet for copy- editing and to improve readability. After using this tool/service, the author reviewed and edited the content as needed and takes full responsibility for the content of the publication.
\printbibliography

@inproceedings{durmus-cardie-2019-corpus,
    title = "A Corpus for Modeling User and Language Effects in Argumentation on Online Debating",
    author = "Durmus, Esin  and
      Cardie, Claire",
    editor = "Korhonen, Anna  and
      Traum, David  and
      M{\`a}rquez, Llu{\'i}s",
    booktitle = "Proceedings of the 57th Annual Meeting of the Association for Computational Linguistics",
    month = jul,
    year = "2019",
    address = "Florence, Italy",
    publisher = "Association for Computational Linguistics",
    url = "https://aclanthology.org/P19-1057/",
    doi = "10.18653/v1/P19-1057",
    pages = "602--607"
}

@inproceedings{durmus-cardie-2018-exploring,
    title = "Exploring the Role of Prior Beliefs for Argument Persuasion",
    author = "Durmus, Esin  and
      Cardie, Claire",
    editor = "Walker, Marilyn  and
      Ji, Heng  and
      Stent, Amanda",
    booktitle = "Proceedings of the 2018 Conference of the North {A}merican Chapter of the Association for Computational Linguistics: Human Language Technologies, Volume 1 (Long Papers)",
    month = jun,
    year = "2018",
    address = "New Orleans, Louisiana",
    publisher = "Association for Computational Linguistics",
    url = "https://aclanthology.org/N18-1094/",
    doi = "10.18653/v1/N18-1094",
    pages = "1035--1045"
}

@article{barbera2015birds, 
title={Birds of the Same Feather Tweet Together: Bayesian Ideal Point Estimation Using Twitter Data}, 
volume={23}, 
DOI={10.1093/pan/mpu011}, 
number={1}, 
journal={Political Analysis}, 
author={Barberá, Pablo}, 
year={2015}, 
pages={76–91}}

@article{barbera2015tweeting,
author = {Pablo Barberá and John T. Jost and Jonathan Nagler and Joshua A. Tucker and Richard Bonneau},
title ={Tweeting From Left to Right: Is Online Political Communication More Than an Echo Chamber?},

journal = {Psychological Science},
volume = {26},
number = {10},
pages = {1531-1542},
year = {2015},
doi = {10.1177/0956797615594620},

URL = { 
    
        https://doi.org/10.1177/0956797615594620
    
    

},
eprint = { 
    
        https://doi.org/10.1177/0956797615594620
    
    

}
,

}

@misc{stan,
  author = {{Stan Development Team}},
  title = {Stan Modeling Language Users Guide and Reference Manual},
  year = {2026},
  url = {https://mc-stan.org},
}

@incollection{betancourt2015hamiltonian,
  author    = {Betancourt, Michael and Girolami, Mark},
  title     = {Hamiltonian Monte Carlo for Hierarchical Models},
  booktitle = {Current Trends in Bayesian Methodology with Applications},
  editor     = {Upadhyay, Satyanshu K. and Singh, Umesh and
                Dey, Dipak K. and Loganathan, Appaia},
  publisher  = {Chapman and Hall/CRC},
  year       = {2015},
  pages      = {79--101},
  doi        = {10.1201/b18502}
}

@inproceedings{reimers2019sentence,
    title = "Sentence-{BERT}: Sentence Embeddings using {S}iamese {BERT}-Networks",
    author = "Reimers, Nils  and
      Gurevych, Iryna",
    editor = "Inui, Kentaro  and
      Jiang, Jing  and
      Ng, Vincent  and
      Wan, Xiaojun",
    booktitle = "Proceedings of the 2019 Conference on Empirical Methods in Natural Language Processing and the 9th International Joint Conference on Natural Language Processing (EMNLP-IJCNLP)",
    month = nov,
    year = "2019",
    address = "Hong Kong, China",
    publisher = "Association for Computational Linguistics",
    url = "https://aclanthology.org/D19-1410/",
    doi = "10.18653/v1/D19-1410",
    pages = "3982--3992"
}

@article{clinton2004statistical,
title={The Statistical Analysis of Roll Call Data}, 
volume={98}, 
DOI={10.1017/S0003055404001194}, 
number={2}, 
journal={American Political Science Review}, 
author={Clinton, Joshua and Jackman, Simon and Rivers, Douglas}, 
year={2004}, 
pages={355–370}}

@article{jackman2001multidimensional,
 title={Multidimensional Analysis of Roll Call Data via Bayesian Simulation: Identification, Estimation, Inference, and Model Checking}, 
 volume={9}, 
 DOI={10.1093/polana/9.3.227}, 
 number={3}, 
 journal={Political Analysis}, 
 author={Jackman, Simon}, 
 year={2001}, 
 pages={227–241}}

@article{poole1985spatial,
 ISSN = {00925853, 15405907},
 URL = {http://www.jstor.org/stable/2111172},
 author = {Keith T. Poole and Howard Rosenthal},
 journal = {American Journal of Political Science},
 number = {2},
 pages = {357--384},
 publisher = {[Midwest Political Science Association, Wiley]},
 title = {A Spatial Model for Legislative Roll Call Analysis},
 urldate = {2026-04-08},
 volume = {29},
 year = {1985}
}

@article{skitka2010psychology,
author = {Skitka, Linda J.},
title = {The Psychology of Moral Conviction},
journal = {Social and Personality Psychology Compass},
volume = {4},
number = {4},
pages = {267-281},
doi = {https://doi.org/10.1111/j.1751-9004.2010.00254.x},
url = {https://compass.onlinelibrary.wiley.com/doi/abs/10.1111/j.1751-9004.2010.00254.x},
eprint = {https://compass.onlinelibrary.wiley.com/doi/pdf/10.1111/j.1751-9004.2010.00254.x},
year = {2010}
}

@misc{anesGuide,
  author = {{American National Election Studies}},
  title  = {The ANES Guide to Public Opinion and Electoral Behavior},
  year   = {2026},
  url    = {https://electionstudies.org/data-tools/anes-guide/},
  note   = {Accessed: 2026-06-01}
}

@article{muchnik2013social,
author = {Lev Muchnik  and Sinan Aral  and Sean J. Taylor },
title = {Social Influence Bias: A Randomized Experiment},
journal = {Science},
volume = {341},
number = {6146},
pages = {647-651},
year = {2013},
doi = {10.1126/science.1240466},
URL = {https://www.science.org/doi/abs/10.1126/science.1240466},
eprint = {https://www.science.org/doi/pdf/10.1126/science.1240466}}

@article{salganik2006experimental,
author = {Matthew J. Salganik  and Peter Sheridan Dodds  and Duncan J. Watts },
title = {Experimental Study of Inequality and Unpredictability in an Artificial Cultural Market},
journal = {Science},
volume = {311},
number = {5762},
pages = {854-856},
year = {2006},
doi = {10.1126/science.1121066},
URL = {https://www.science.org/doi/abs/10.1126/science.1121066},
eprint = {https://www.science.org/doi/pdf/10.1126/science.1121066},
}

@article{chaiken1980heuristic,
  title = {Heuristic versus systematic information processing and the use of source versus message cues in persuasion.},
  author = {Chaiken, Shelly},
  journal = {Journal of personality and social psychology},
  volume = {39},
  number = {5},
  pages = {752-766},
  year = {1980},
  publisher = {American Psychological Association},
  doi = {10.1037/0022-3514.39.5.752},
}

@book{sunstein2009going,
  title = {Going to extremes: How like minds unite and divide},
  author = {Sunstein, Cass R},
  year = {2009},
  publisher = {Oxford University Press},
  doi = {10.1093/oso/9780195378016.001.0001}
}

@article{lorenz2011social,
author = {Jan Lorenz  and Heiko Rauhut  and Frank Schweitzer  and Dirk Helbing },
title = {How social influence can undermine the wisdom of crowd effect},
journal = {Proceedings of the National Academy of Sciences},
volume = {108},
number = {22},
pages = {9020-9025},
year = {2011},
doi = {10.1073/pnas.1008636108},
URL = {https://www.pnas.org/doi/abs/10.1073/pnas.1008636108},
eprint = {https://www.pnas.org/doi/pdf/10.1073/pnas.1008636108}
}

@inproceedings{adamic2005political,
author = {Adamic, Lada A. and Glance, Natalie},
title = {The political blogosphere and the 2004 U.S. election: divided they blog},
year = {2005},
isbn = {1595932151},
publisher = {Association for Computing Machinery},
address = {New York, NY, USA},
url = {https://doi.org/10.1145/1134271.1134277},
doi = {10.1145/1134271.1134277},
booktitle = {Proceedings of the 3rd International Workshop on Link Discovery},
pages = {36–43},
numpages = {8},
location = {Chicago, Illinois},
series = {LinkKDD '05}
}

@article{flamino2023political,
  title = {Political polarization of news media and influencers on Twitter in the 2016 and 2020 US presidential elections},
  author = {Flamino, James and Galeazzi, Alessandro and Feldman, Stuart and Macy, Michael W and Cross, Brendan and Zhou, Zhenkun and Serafino, Matteo and Bovet, Alexandre and Makse, Hern{\'a}n A and Szymanski, Boleslaw K},
  journal = {Nature Human Behaviour},
  volume = {7},
  pages = {904--916},
  year = {2023},
  publisher = {Nature Publishing Group UK London},
  doi = {10.1038/s41562-023-01550-8},
}

@article{tornberg2022digital,
author = {Petter Törnberg },
title = {How digital media drive affective polarization through partisan sorting},
journal = {Proceedings of the National Academy of Sciences},
volume = {119},
number = {42},
pages = {e2207159119},
year = {2022},
doi = {10.1073/pnas.2207159119},
URL = {https://www.pnas.org/doi/abs/10.1073/pnas.2207159119},
eprint = {https://www.pnas.org/doi/pdf/10.1073/pnas.2207159119},
}

@article{skitka2021psychology,
   author = "Skitka, Linda J. and Hanson, Brittany E. and Morgan, G. Scott and Wisneski, Daniel C.",
   title = "The Psychology of Moral Conviction", 
   journal= "Annual Review of Psychology",
   year = "2021",
   volume = "72",
   number = "Volume 72, 2021",
   pages = "347-366",
   doi = "https://doi.org/10.1146/annurev-psych-063020-030612",
   url = "https://www.annualreviews.org/content/journals/10.1146/annurev-psych-063020-030612",
   publisher = "Annual Reviews",
   issn = "1545-2085",
   type = "Journal Article",
  }

@book{baker2004item,
  title = {Item response theory: Parameter estimation techniques},
  author = {Baker, Frank B and Kim, Seock-Ho},
  year = {2004},
  publisher = {CRC press},
  doi = {10.1201/9781482276725},
}

@article{aramovich2012opposing,
author = {Nicholas P. Aramovich and Brad L. Lytle and Linda J. Skitka},
title = {Opposing torture: Moral conviction and resistance to majority influence},
journal = {Social Influence},
volume = {7},
number = {1},
pages = {21--34},
year = {2012},
publisher = {Routledge},
doi = {10.1080/15534510.2011.640199},


URL = { 
    
        https://doi.org/10.1080/15534510.2011.640199
    
    

},
eprint = { 
    
        https://doi.org/10.1080/15534510.2011.640199
    
    

}

}

@article{silva2024robust,
author = {Luca Alessandro Silva and Giacomo Zanella},
title = {Robust Leave-One-Out Cross-Validation for High-Dimensional Bayesian Models},
journal = {Journal of the American Statistical Association},
volume = {119},
number = {547},
pages = {2369--2381},
year = {2024},
publisher = {Taylor \& Francis},
doi = {10.1080/01621459.2023.2257893},


URL = { 
    
        https://doi.org/10.1080/01621459.2023.2257893
    
    

},
eprint = { 
    
        https://doi.org/10.1080/01621459.2023.2257893
    
    

}

}

@article{lee2025semantic,
  title = {A semantic embedding space based on large language models for modelling human beliefs},
  author = {Lee, Byunghwee and Aiyappa, Rachith and Ahn, Yong-Yeol and Kwak, Haewoon and An, Jisun},
  journal = {Nature Human Behaviour},
  volume = {9},
  number = {9},
  pages = {1928--1940},
  year = {2025},
  publisher = {Nature Publishing Group UK London},
  doi = {10.1038/s41562-025-02228-z},
}

@article{lee2026llms,
  title = {LLMs Can Infer Political Alignment from Online Conversations},
  author = {Lee, Byunghwee and Kim, Sangyeon and Menczer, Filippo and Ahn, Yong-Yeol and Kwak, Haewoon and An, Jisun},
  journal = {arXiv preprint arXiv:2603.11253},
  year = {2026},
}

@misc{qixiang2026,
author = {Fang, Qixiang and Garcia-Bernardo, Javier and Kesteren, Erik-Jan},
year = {2026},
month = {03},
pages = {},
title = {A Methodological Guide on Using Large Language Models for Text Annotation in the Social Sciences and Humanities with Python and R},
doi = {10.31235/osf.io/v4eq6_v2}
}

@misc{seckin2026emergence,
      title={Emergence of Stereotypes and Affective Polarization from Belief Network Dynamics}, 
      author={Ozgur Can Seckin and Rachith Aiyappa and Madalina Vlasceanu and Filippo Menczer and Alessandro Flammini and Yong-Yeol Ahn},
      year={2026},
      eprint={2604.10251},
      archivePrefix={arXiv},
      primaryClass={cs.SI},
      url={https://arxiv.org/abs/2604.10251}, 
      doi={10.48550/arXiv.2604.10251}
}

@article{Kunda1990,
  author  = {Kunda, Ziva},
  title   = {The Case for Motivated Reasoning},
  journal = {Psychological Bulletin},
  year    = {1990},
  volume  = {108},
  number  = {3},
  pages   = {480--498},
  doi     = {10.1037/0033-2909.108.3.480},
}

@misc{Priniski2022,
  author       = {Priniski, J. and Solanki, Prachi and Horne, Zachary},
  title        = {A {B}ayesian Decision-Theoretic Framework for Studying Motivated Reasoning},
  year         = {2022},
  publisher    = {OSF},
  howpublished = {OSF Preprints},
}

@book{maher1993betting,
  title={Betting on theories},
  author={Maher, Patrick},
  year={1993},
  publisher={Cambridge University Press}
}

@article{Rigoli19012026,
author = {Francesco Rigoli},
title = {Bayes or Pascal? The computations underlying motivated reasoning},
journal = {Philosophical Psychology},
volume = {0},
number = {0},
pages = {1--36},
year = {2026},
publisher = {Routledge},
doi = {10.1080/09515089.2026.2615690},


URL = { 
    
        https://doi.org/10.1080/09515089.2026.2615690
    
    

},
eprint = { 
    
        https://doi.org/10.1080/09515089.2026.2615690
    
    

}

}

@article{
Horne2015,
author = {Zachary Horne  and Derek Powell  and John E. Hummel  and Keith J. Holyoak },
title = {Countering antivaccination attitudes},
journal = {Proceedings of the National Academy of Sciences},
volume = {112},
number = {33},
pages = {10321-10324},
year = {2015},
doi = {10.1073/pnas.1504019112},
URL = {https://www.pnas.org/doi/abs/10.1073/pnas.1504019112},
eprint = {https://www.pnas.org/doi/pdf/10.1073/pnas.1504019112}}

@article{Powell2023,
  author  = {Powell, Derek and Weisman, Kara and Markman, Ellen M.},
  title   = {Modeling and Leveraging Intuitive Theories to Improve Vaccine Attitudes},
  journal = {Journal of Experimental Psychology: General},
  year    = {2023},
  volume  = {152},
  number  = {5},
  pages   = {1379--1395},
  doi     = {10.1037/xge0001324},
}

@article{manski1993,
 ISSN = {00346527, 1467937X},
 URL = {http://www.jstor.org/stable/2298123},
 author = {Charles F. Manski},
 journal = {The Review of Economic Studies},
 number = {3},
 pages = {531--542},
 publisher = {[Oxford University Press, Review of Economic Studies, Ltd.]},
 title = {Identification of Endogenous Social Effects: The Reflection Problem},
 urldate = {2026-08-03},
 volume = {60},
 year = {1993}
}

@article{bail2018exposure,
author = {Christopher A. Bail  and Lisa P. Argyle  and Taylor W. Brown  and John P. Bumpus  and Haohan Chen  and M. B. Fallin Hunzaker  and Jaemin Lee  and Marcus Mann  and Friedolin Merhout  and Alexander Volfovsky },
title = {Exposure to opposing views on social media can increase political polarization},
journal = {Proceedings of the National Academy of Sciences},
volume = {115},
number = {37},
pages = {9216-9221},
year = {2018},
doi = {10.1073/pnas.1804840115},
URL = {https://www.pnas.org/doi/abs/10.1073/pnas.1804840115},
eprint = {https://www.pnas.org/doi/pdf/10.1073/pnas.1804840115}}

@book{mutz2006hearing,
  title={Hearing the other side: Deliberative versus participatory democracy},
  author={Mutz, Diana C},
  year={2006},
  publisher={Cambridge University Press},
  doi={https://doi.org/10.1017/CBO9780511617201}
}

@inbook{chambers2025deliberation, place={Cambridge}, series={Cambridge Law Handbooks}, title={Deliberation}, booktitle={The Cambridge Handbook of Constitutional Theory}, publisher={Cambridge University Press}, author={Chambers, Simone}, editor={Bellamy, Richard and King, JeffEditors}, year={2025}, pages={246–261}, collection={Cambridge Law Handbooks}}

@book{mouffe2000democratic,
  author    = {Mouffe, Chantal},
  title     = {The Democratic Paradox},
  publisher = {Verso},
  year      = {2000},
  isbn      = {9781859842799}
}

@article{hoffman2014nuts,
author = {Hoffman, Matthew D. and Gelman, Andrew},
title = {The No-U-turn sampler: adaptively setting path lengths in Hamiltonian Monte Carlo},
year = {2014},
issue_date = {January 2014},
publisher = {JMLR.org},
volume = {15},
number = {1},
issn = {1532-4435},
journal = {J. Mach. Learn. Res.},
month = jan,
pages = {1593–1623},
numpages = {31}
}

@article{conover2018republicans,
  title={How republicans won on voter identification laws: The roles of strategic reasoning and moral conviction},
  author={Conover, Pamela Johnston and Miller, Patrick R},
  journal={Social Science Quarterly},
  volume={99},
  number={2},
  pages={490--511},
  year={2018},
  publisher={Wiley Online Library}
}

@book{mooney2001clash,
  title = {The Public Clash of Private Values: {{The}} Politics of Morality Policy},
  author = {Mooney, C.Z.},
  date = {2001},
  publisher = {SAGE Publications},
  url = {https://books.google.nl/books?id=npmAAAAAIAAJ},
  isbn = {978-1-889119-40-3},
  lccn = {99507433}
}

@incollection{petty1986elaboration,
title = {The Elaboration Likelihood Model of Persuasion},
editor = {Leonard Berkowitz},
series = {Advances in Experimental Social Psychology},
publisher = {Academic Press},
volume = {19},
pages = {123-205},
year = {1986},
issn = {0065-2601},
doi = {https://doi.org/10.1016/S0065-2601(08)60214-2},
url = {https://www.sciencedirect.com/science/article/pii/S0065260108602142},
author = {Richard E. Petty and John T. Cacioppo}
}

@inproceedings{habernal2016convincing,
    title = "What makes a convincing argument? Empirical analysis and detecting attributes of convincingness in Web argumentation",
    author = "Habernal, Ivan  and
      Gurevych, Iryna",
    editor = "Su, Jian  and
      Duh, Kevin  and
      Carreras, Xavier",
    booktitle = "Proceedings of the 2016 Conference on Empirical Methods in Natural Language Processing",
    month = nov,
    year = "2016",
    address = "Austin, Texas",
    publisher = "Association for Computational Linguistics",
    url = "https://aclanthology.org/D16-1129/",
    doi = "10.18653/v1/D16-1129",
    pages = "1214--1223"
}

@article{esau2021arenas,
author = {Esau, Katharina and Fleuß, Dannica and Nienhaus, Sarah-Michelle},
title = {Different Arenas, Different Deliberative Quality? Using a Systemic Framework to Evaluate Online Deliberation on Immigration Policy in Germany},
journal = {Policy \& Internet},
volume = {13},
number = {1},
pages = {86-112},
doi = {https://doi.org/10.1002/poi3.232},
url = {https://onlinelibrary.wiley.com/doi/abs/10.1002/poi3.232},
eprint = {https://onlinelibrary.wiley.com/doi/pdf/10.1002/poi3.232},
year = {2021}
}

@inproceedings{tan2016winning,
author = {Tan, Chenhao and Niculae, Vlad and Danescu-Niculescu-Mizil, Cristian and Lee, Lillian},
title = {Winning Arguments: Interaction Dynamics and Persuasion Strategies in Good-faith Online Discussions},
year = {2016},
isbn = {9781450341431},
publisher = {International World Wide Web Conferences Steering Committee},
address = {Republic and Canton of Geneva, CHE},
url = {https://doi.org/10.1145/2872427.2883081},
doi = {10.1145/2872427.2883081},
booktitle = {Proceedings of the 25th International Conference on World Wide Web},
pages = {613–624},
numpages = {12},
location = {Montr{\'e}al, Qu{\'e}bec, Canada},
series = {WWW '16}
}

@article{Martin2026arviz,
doi = {10.21105/joss.09889},
url = {https://doi.org/10.21105/joss.09889},
year = {2026},
publisher = {The Open Journal},
volume = {11},
number = {119},
pages = {9889},
author = {Martin, Osvaldo A. and Abril-Pla, Oriol and Deklerk, Jordan and Axen, Seth D. and Carroll, Colin and Hartikainen, Ari and Vehtari, Aki},
title = {ArviZ: a modular and flexible library for exploratory analysis of Bayesian models},
journal = {Journal of Open Source Software}}

@article{bartels2002beyond,
  title={Beyond the running tally: Partisan bias in political perceptions},
  author={Bartels, Larry M},
  journal={Political behavior},
  volume={24},
  number={2},
  pages={117--150},
  year={2002},
  publisher={Springer},
  doi ={10.1023/A:1021226224601}
}

@article{maxey2026moralization,
  title={Moralization and Foreign Policy Attitudes},
  author={Maxey, Sarah and Powers, Kathleen E},
  journal={Political Behavior},
  volume={48},
  number={2},
  pages={1167--1196},
  year={2026},
  publisher={Springer},
  doi={10.1007/s11109-025-10064-0}
}

\clearpage
\appendix
\renewcommand\thefigure{A\arabic{figure}}  
\renewcommand\thetable{A\arabic{table}}  
\setcounter{figure}{0}
\setcounter{table}{0}
\section{Appendix}
\subsection{User demographics and prior beliefs}\label{app:user-demographics}
\begin{figure}
    \centering
    \includegraphics[width=0.9\linewidth]{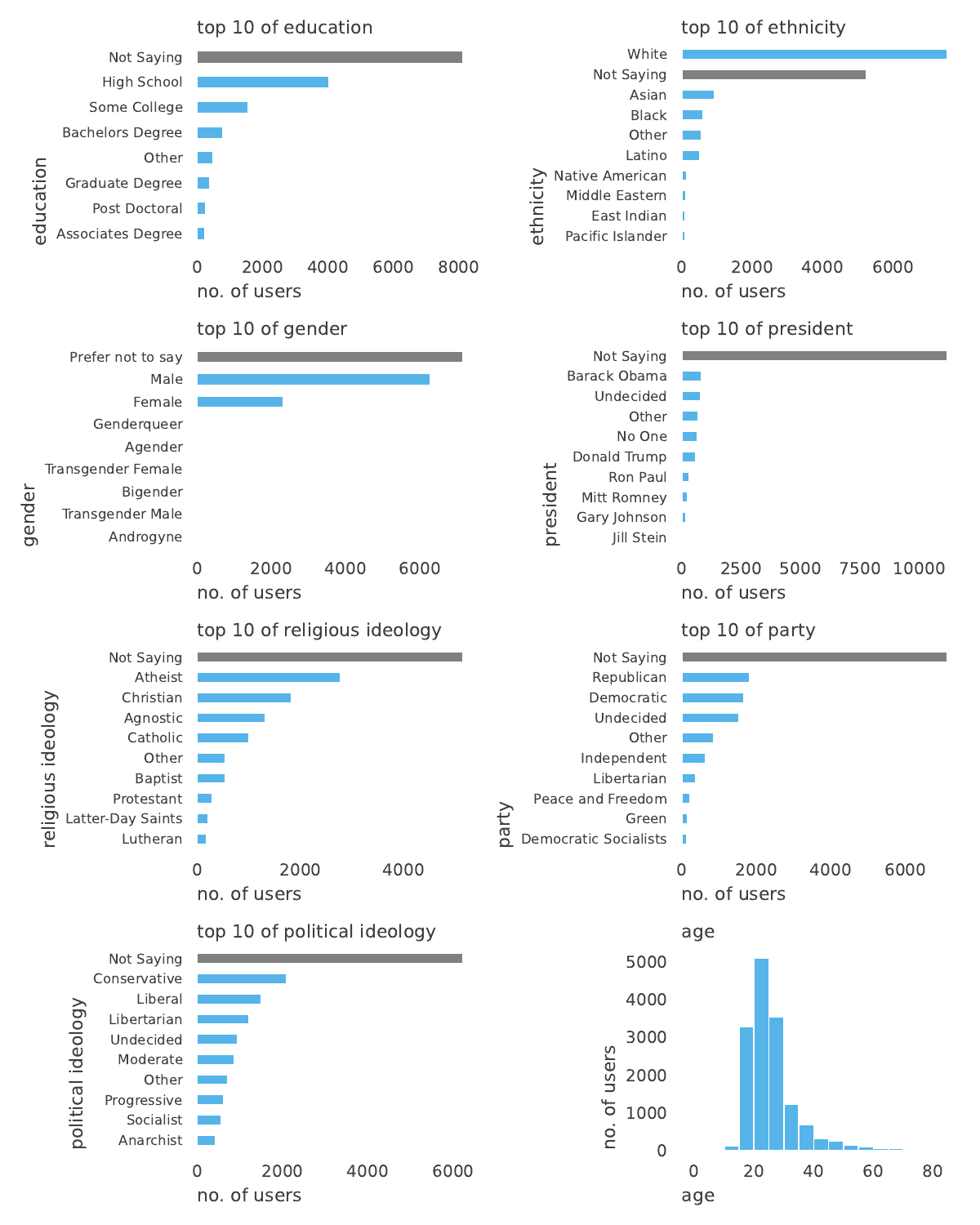}
    \caption{\textbf{Users demographics.} Each bar plot shows the prevalence of each demographic category, respectively: education, ethnicity, gender, favorite U.S. president, religion, party, political ideology, and age.}
    \label{fig:demographics}
\end{figure}

Figure~\ref{fig:demographics} summarises the demographic composition of the 15,840 users retained for analysis. The user base is predominantly young, with the age distribution peaking around 18–25 years and a long right tail extending to older age groups. In terms of gender, the majority of users who disclosed their gender identify as male, with female users representing a smaller but substantial share; a non-negligible fraction preferred not to disclose. The ethnicity distribution is heavily skewed toward users identifying as white, with other ethnic groups representing small minorities. A large proportion of users chose ``Not Saying'' across several demographic dimensions, reflecting the voluntary nature of self-reporting on the platform.

Politically, the user base spans a broad ideological range. Among those who disclosed a political ideology, Conservative and Liberal are the two most represented categories, followed by Libertarian and smaller groups including Moderate, Progressive, and Socialist. Party affiliation broadly mirrors this pattern, with Republican and Democratic users being most prevalent, alongside a notable share of ``Undecided'' and Independent users. Regarding religion, Atheist and Christian are the most common self-reported categories, followed by Agnostic and Catholic, reflecting a user base that skews secular relative to the general U.S. population. On presidential preference, Barack Obama is the most frequently cited favourite president among users who disclosed a preference, followed at a distance by ``Undecided'' and a variety of other figures, including Ron Paul and Donald Trump, consistent with the platform's active period spanning the Obama and early Trump eras.

These demographics suggest that the \textit{debate.org} user base, while diverse in political ideology, is not representative of the broader U.S. or global population. Users tend to be young, male, white, and politically engaged, which should be borne in mind when interpreting the generalisability of the results.
\clearpage
\subsection{Prior beliefs}
\begin{figure}[htbp]
    \centering
    \includegraphics[width=0.8\linewidth]{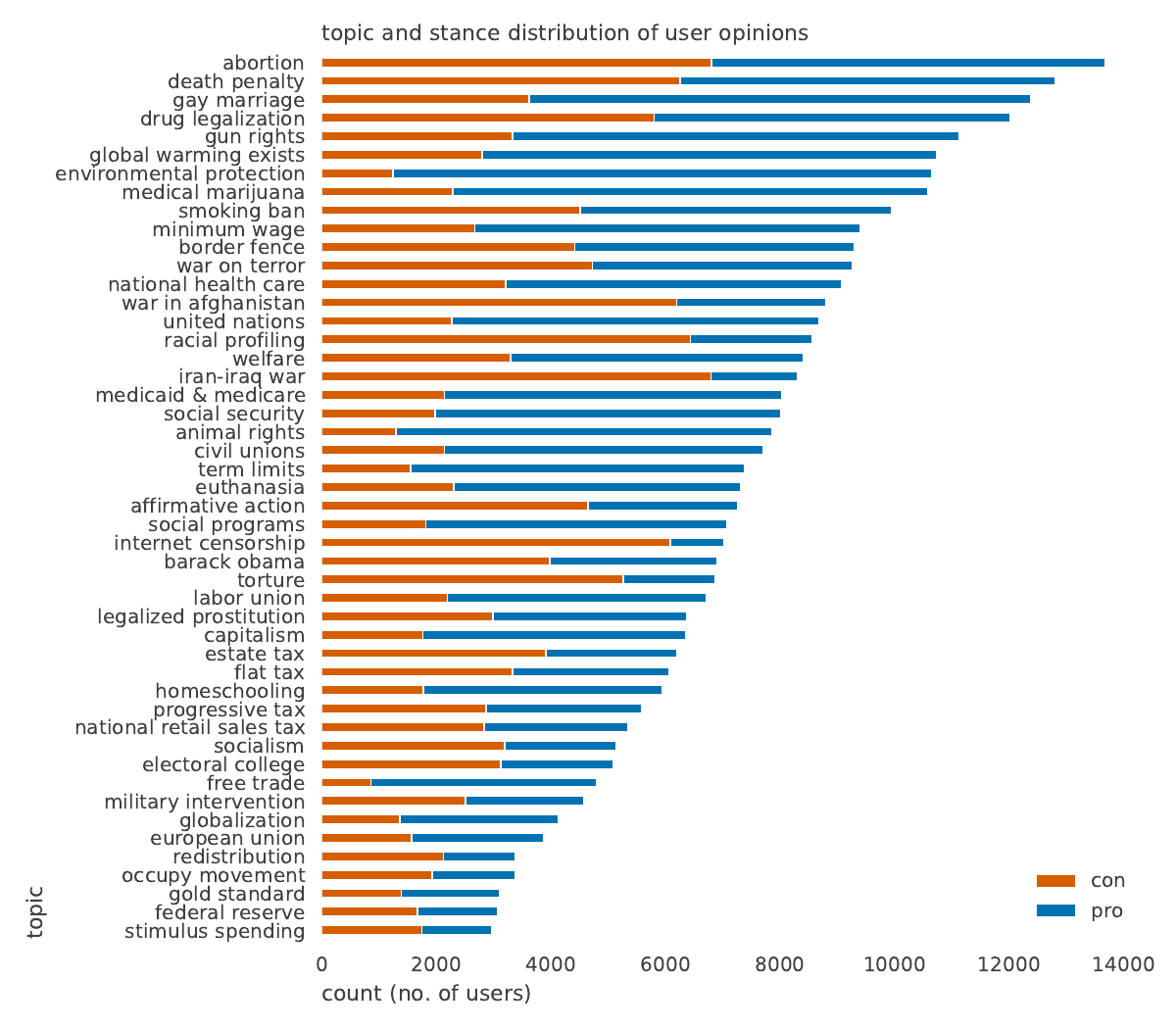}
    \caption{\textbf{User prior beliefs.} Topic-wise distributions of
    PRO/CON/NEUTRAL counts.}
    \label{fig:users-prior-beliefs}
\end{figure}

Figure~\ref{fig:users-prior-beliefs} shows the distribution of self-reported prior beliefs across the 48 topics, separately for \textbf{PRO} and \textbf{CON} stances. The most frequently reported topics, abortion, death penalty, gay marriage, drug legalisation, and gun rights, are also among the most debated on the platform (see Figures~\ref{fig:debate-class-gpt} and \ref{fig:debate-class-bert}), suggesting a coherent relationship between users' expressed interests and their engagement with debates.
For most topics, \textbf{PRO} stances are more common than \textbf{CON} stances, reflecting either a genuine asymmetry in user opinion or a response bias in self-reporting. The asymmetry is particularly pronounced for topics such as environmental protection, global warming exists, gay marriage, where \textbf{PRO} responses substantially outnumber \textbf{CON} responses. The reverse pattern, where \textbf{CON} responses are more common, is rarer but visible for topics such as abortion, where the split is closer, and for racial profiling, where \textbf{CON} responses predominate.
Topics at the bottom of the distribution, such as stimulus spending, federal reserve, gold standard, and occupy movement, have relatively few users reporting any stance, indicating lower salience or familiarity among the platform's user base. This uneven coverage across topics is relevant to the interpretation of the model results: topics with sparse prior belief data yield noisier topic-alignment estimates and wider credible intervals for the corresponding coefficients.
\clearpage

\subsection{Debate classification robustness}\label{app:debate-classification}
The debate annotation procedure using LLMs is inspired by the methodological guide by Fang et al~\cite{qixiang2026}. I use the following prompt for GPT-4o-mini:
\begin{tcolorbox}[breakable, colback=white, colframe=black, boxrule=0.8pt]
\begin{verbatim}
You are a political debate analyzer. 
You will be given a debate title and participants' comments.

## TASKS ##
1. Identify the single most appropriate debate topic based on the debate title 
and all participant comments.
2. Classify the stance of the debate with respect to the identified topic.

## STANCE DEFINITION ##
The stance must be evaluated **relative to the identified topic**, 
not relative to the specific policy, object, or practice mentioned in the debate title.

- "pro": The debate supports, promotes, or advances the values, goals, or principles 
         of the identified topic.
- "con": The debate opposes, criticizes, or undermines the values, goals, or principles 
         of the identified topic.
- "neutral": The debate is descriptive, mixed, or unclear with 
             respect to the identified topic.

IMPORTANT:
- If a debate argues against a practice because it conflicts with the values of a topic, 
the stance should be "pro" toward that topic.
  (Example: "Zoos should be banned" → topic: "animal rights" → stance: "pro")

## OUTPUT FORMAT ##
Return the output in the following JSON format and nothing else:

{{
  "topic": "<topic_label>",
  "stance": "<stance_label>"
}}

## TOPIC CONSTRAINT ##
Choose ONLY ONE topic from the following list:

['abortion', 'affirmative action', 'animal rights', 'barack obama', 'border fence', 
'capitalism', 'civil unions', 'death penalty', 'drug legalization', 'electoral college', 
'environmental protection','estate tax', 'european union', 'euthanasia', 'federal reserve', 
'flat tax', 'free trade', 'gay marriage', 'global warming exists', 'globalization', 
'gold standard', 'gun rights', 'homeschooling', 'internet censorship', 'iran-iraq war', 
'labor union', 'legalized prostitution', 'medicaid & medicare', 'medical marijuana', 
'military intervention', 'minimum wage', 'national health care', 
'national retail sales tax', 'occupy movement', 'progressive tax', 'racial profiling',
'redistribution', 'smoking ban', 'social programs', 'social security', 'socialism', 
'stimulus spending', 'term limits', 'torture', 'united nations', 'war in afghanistan', 
'war on terror', 'welfare', 'other']
\end{verbatim}
\end{tcolorbox}

\begin{figure}[htbp]
    \centering
    \includegraphics[width=0.8\linewidth]{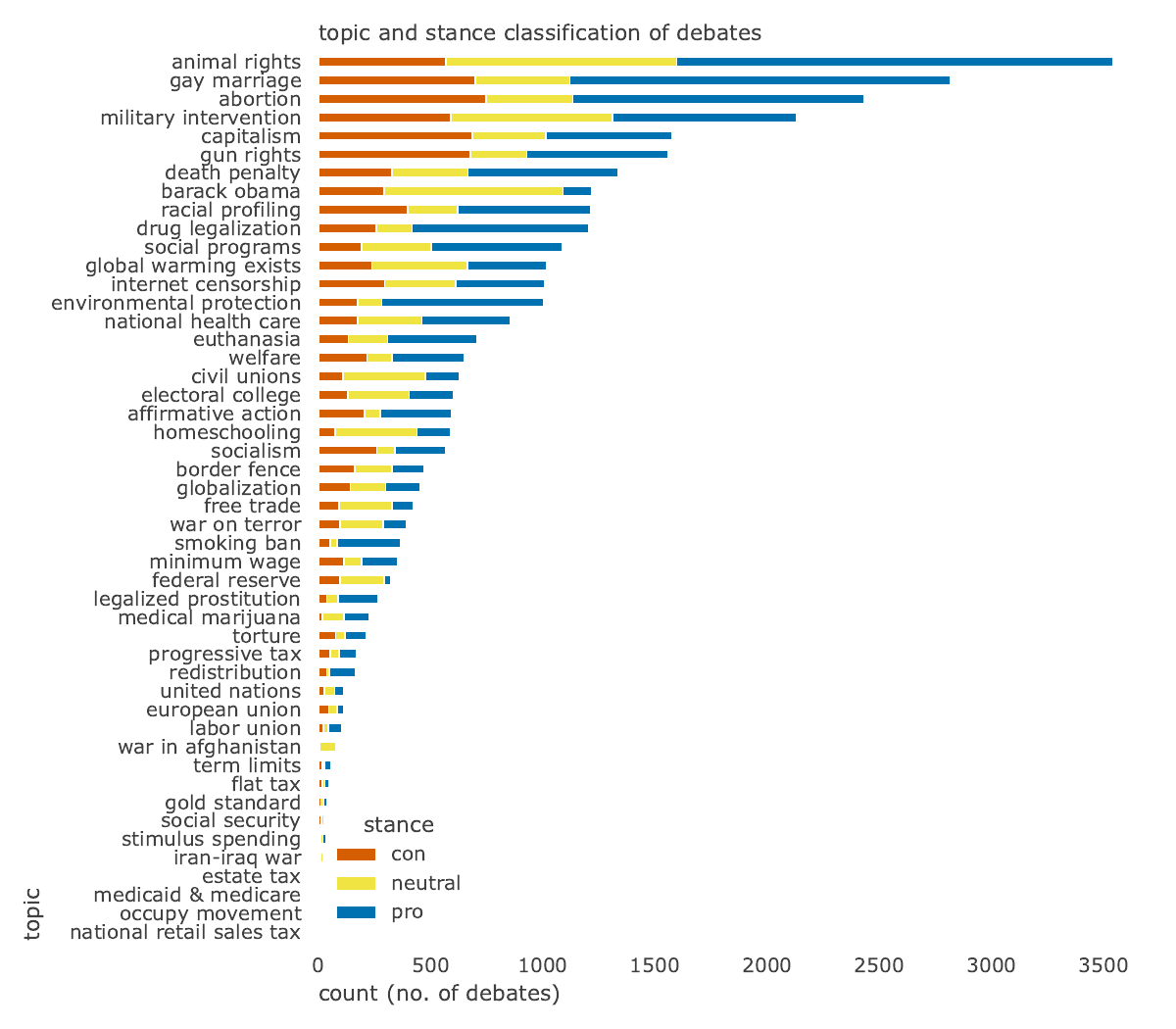}
    \caption{\textbf{Debates stance classifications (GPT-4o-mini).}
    Topic-wise distributions of PRO/CON/NEUTRAL counts.}
    \label{fig:debate-class-gpt}
\end{figure}

\begin{figure}[htbp]
    \centering
    \includegraphics[width=0.8\linewidth]{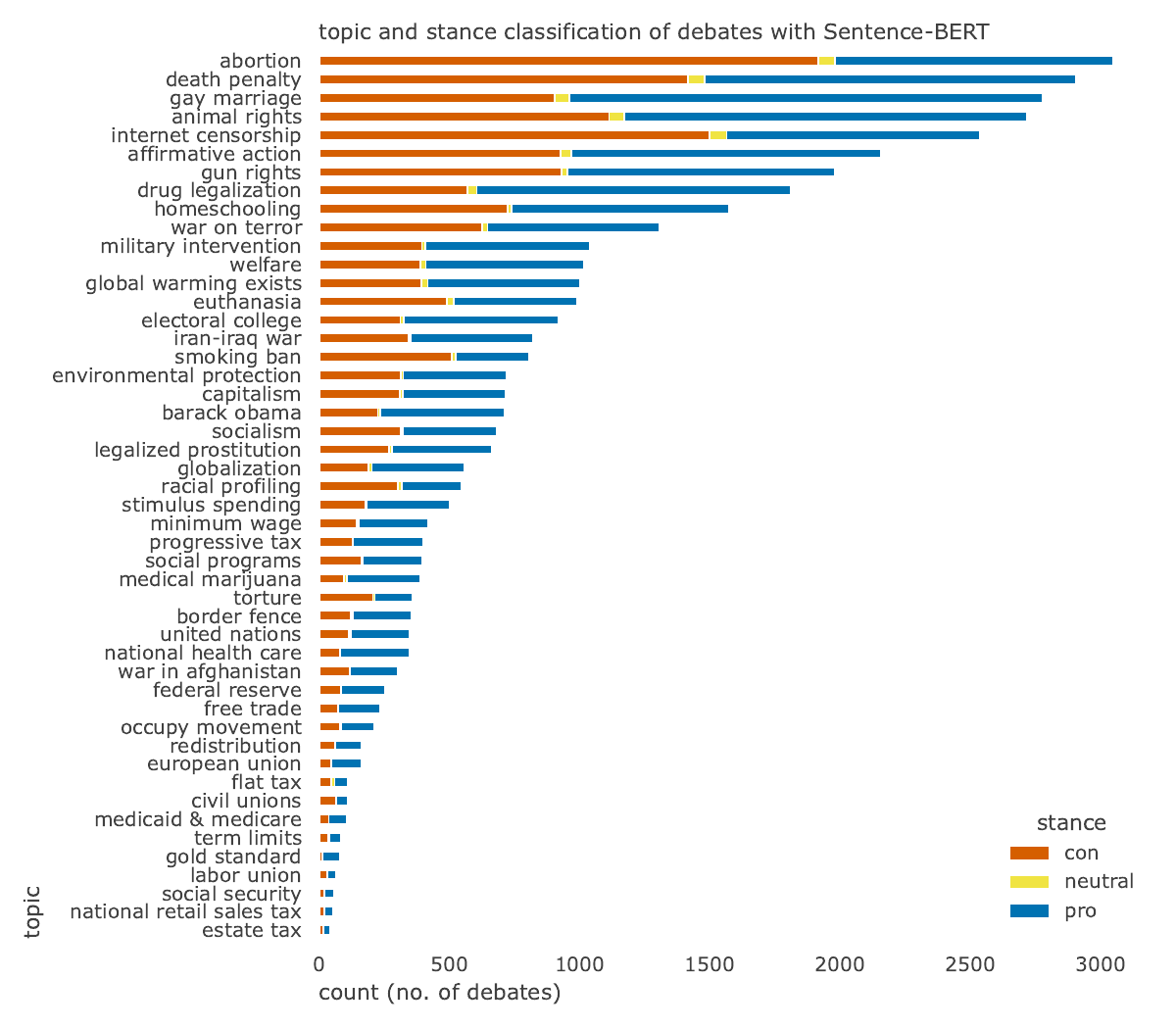}
    \caption{\textbf{Debates stance classifications (SBERT + BART-MNLI).}
    Topic-wise distributions of PRO/CON/NEUTRAL counts.}
    \label{fig:debate-class-bert}
\end{figure}

Figures~\ref{fig:debate-class-gpt} and~\ref{fig:debate-class-bert} show the distribution of debates across the 48 topics classified by the two methods, respectively, GPT-4o-mini and SBERT + BART-MNLI. The inter-coder agreement at the debate level is Krippendorff's $\alpha = 0.55$, indicating low-to-moderate agreement on the topic annotation of individual debates.

\begin{figure}[htbp]
    \centering
    \includegraphics[width=\linewidth]{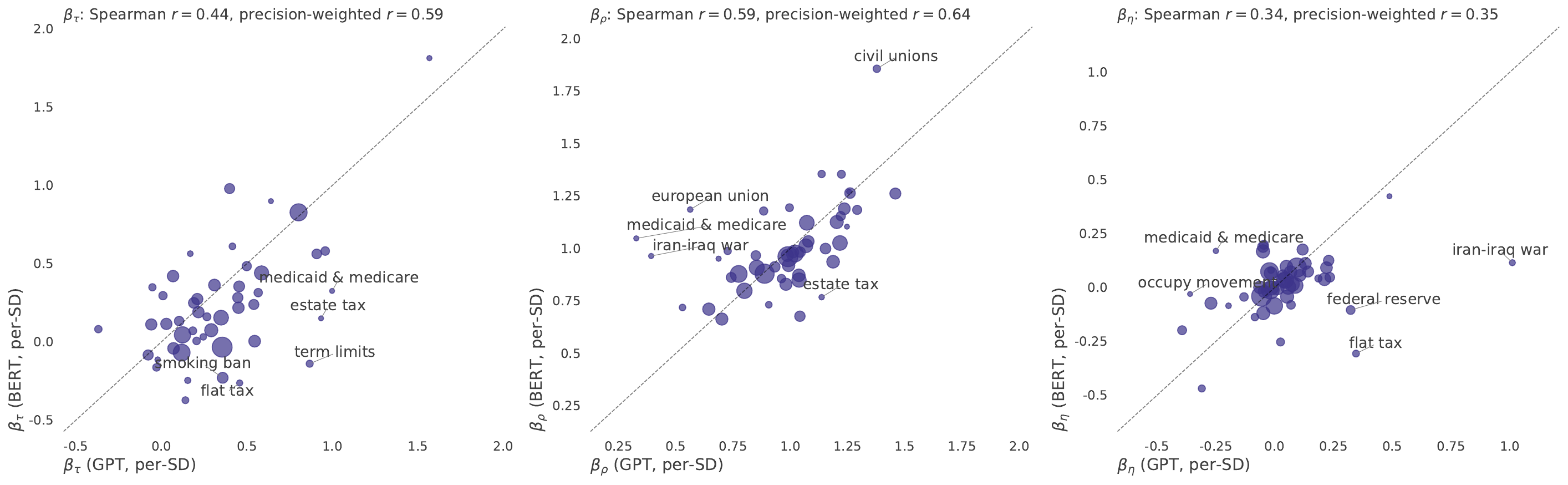}
    \caption{\textbf{Comparison of topic-specific posterior means across
    GPT-4o-mini and SBERT + BART-MNLI debate classifications.} Each point
    represents one topic; point size is proportional to posterior precision
    (inverse CI width), downweighting topics estimated imprecisely under either
    classifier. Left: conviction coefficients $\beta_\conv$ (Spearman
    $r = 0.44$, precision-weighted $r = 0.59$). Middle: persuasiveness
    coefficients $\beta_\pers$ (Spearman $r = 0.59$, precision-weighted $r = 0.64$).
    Right: cross-cutting exposure main-effect coefficients $\beta_\expo$ (Spearman
    $r = 0.34$, precision-weighted $r = 0.35$). The dashed line is the identity
    line.}
    \label{fig:robustness-classification}
\end{figure}

To assess the sensitivity of the results to the choice of classification method, I re-estimate the full model using SBERT + BART-MNLI topic and stance classifications in place of GPT-4o-mini, and re-assign RQ1 and RQ2 summary labels (Section~\ref{subsec:met-param-class}) based on the resulting posterior. Posterior means of the topic-specific coefficients are correlated across the two specifications: $r_{\beta_\conv} = 0.44$ (conviction), $r_{\beta_\pers} = 0.59$ (persuasiveness), and $r_{\beta_\expo} = 0.34$ (cross-cutting exposure, main effect). Weighting each Spearman correlation by the harmonic mean of the two classifiers' posterior precision (i.e., downweighting topics estimated imprecisely under either classifier) gives $r_{\beta_\conv} = 0.59$, $r_{\beta_\pers} = 0.64$, and $r_{\beta_\expo} = 0.35$. Figure~\ref{fig:robustness-classification} shows all three comparisons: $\beta_\pers$ is the most stable of the three across classifiers, consistent with it not being constructed from the debate vectors at all, while $\beta_\expo$ is the least stable, consistent with it depending on the full temporal sequence of prior voters' inferred debate topics indirectly (via which debates enter the audience-composition calculation for a given voter).

\begin{figure}[htbp]
    \centering
    \includegraphics[width=0.7\linewidth]{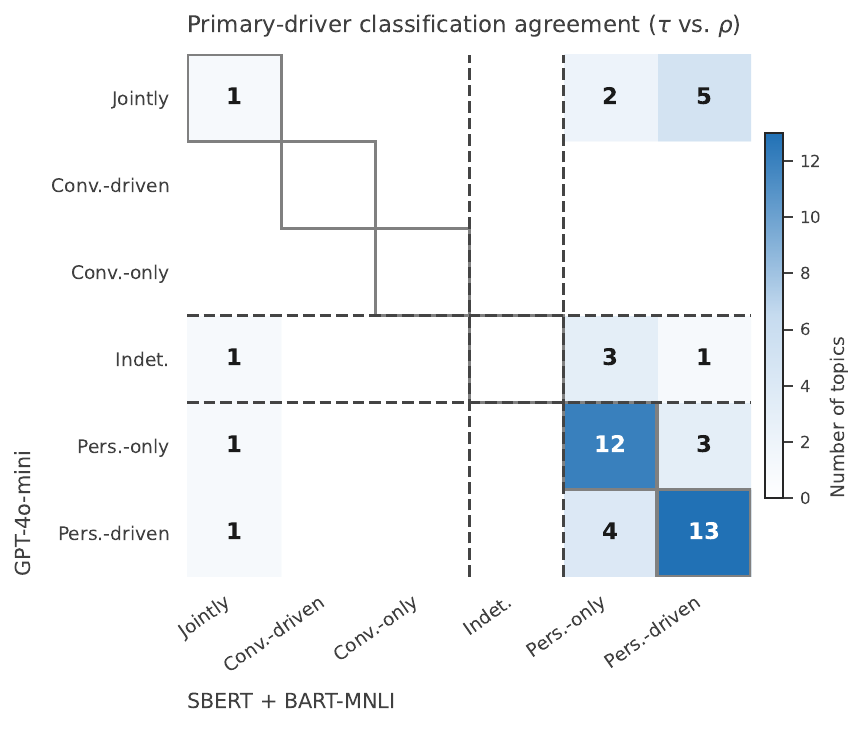}
    \caption{\textbf{Confusion matrix of RQ1 labelling across
    GPT-4o-mini and SBERT + BART-MNLI.} Rows correspond to GPT-4o-mini
    classifications, columns to SBERT + BART-MNLI. Diagonal entries are
    agreements; off-diagonal entries are disagreements. Neither classifier ever
    assigns \emph{conviction-driven} or \emph{conviction-only} to any topic.}
    \label{fig:class-confusion-primary}
\end{figure}

\begin{figure}[htbp]
    \centering
    \includegraphics[width=0.6\linewidth]{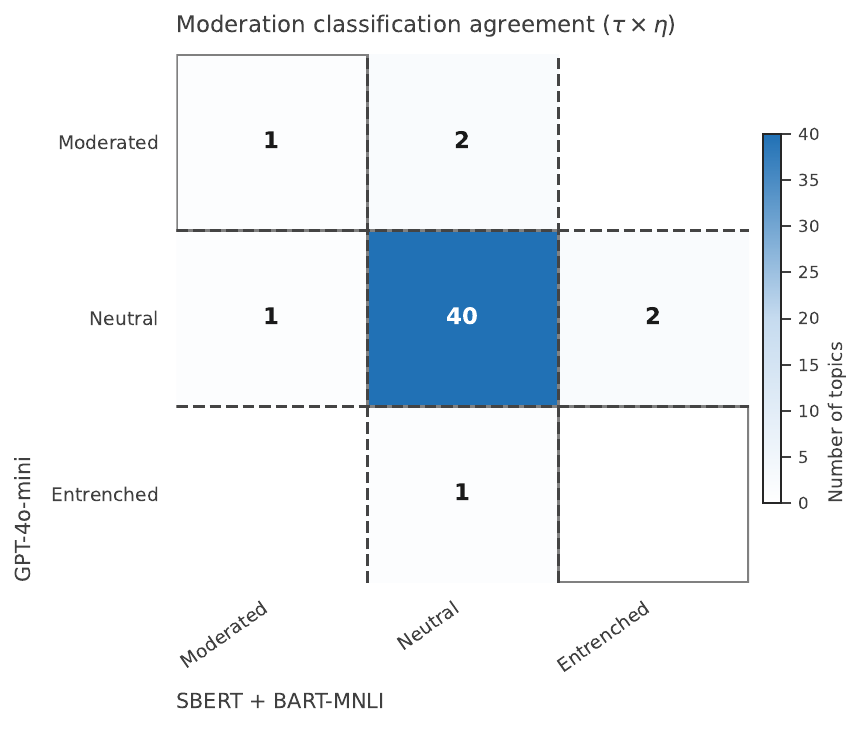}
    \caption{\textbf{Confusion matrix of RQ2 labelling across
    GPT-4o-mini and SBERT + BART-MNLI.} Rows correspond to GPT-4o-mini
    classifications, columns to SBERT + BART-MNLI. Of GPT's three
    \emph{exposure-moderated} topics, only \textit{animal rights} is
    corroborated by BERT; of GPT's one \emph{exposure-entrenched} topic
    (smoking ban), BERT classifies it \emph{exposure-neutral}.}
    \label{fig:class-confusion-moderation}
\end{figure}

Table~\ref{tab:primary-class-distribution} and Table~\ref{tab:moderation-class-distribution} show the distribution of topics across the two classification schemes under each classifier, and Figures~\ref{fig:class-confusion-primary} and~\ref{fig:class-confusion-moderation} show the corresponding confusion matrices, all derived from the posterior estimates of the full model re-estimated under each classifier.

For the \textbf{RQ1} labelling, the two classifiers agree on 26 of 47 topics (55\%), with Krippendorff's $\alpha = 0.31$ and no significant difference in the overall class distributions ($\chi^2 = 5.07$, $df = 3$, $p = 0.17$). One topic present under BERT is absent under GPT's topic set and is dropped from this comparison. Reassuringly, neither classifier ever assigns \emph{conviction-driven} or \emph{conviction-only} to any topic: the finding that conviction never credibly outweighs persuasiveness (Section~\ref{subsec:primary-drivers}) holds under both classifiers, not just GPT's. Every disagreement instead falls along the \emph{strength} axis of the same ordering, i.e., how strong the conviction and persuasiveness signals are judged to be, not which one wins: topics shift between \emph{persuasiveness-only}, \emph{persuasiveness-driven}, \emph{jointly-driven}, and \emph{indeterminate} depending on classifier (e.g., abortion and global warming exists are persuasiveness-only under BERT but persuasiveness-driven under GPT; legalized prostitution, medical marijuana, and minimum wage are persuasiveness-driven under BERT but jointly-driven under GPT). This is consistent with the two classifiers producing correlated but noisy estimates of the same two underlying strengths, rather than disagreeing about the substantive picture.

For the \textbf{RQ2} labelling, agreement is higher: 41 of 47 topics (87\%), with Krippendorff's $\alpha = 0.21$ and no significant difference in class distributions ($\chi^2 = 0.53$, $df = 2$, $p = 0.77$), reflecting the highly imbalanced distribution dominated by the exposure-neutral label. Of the four topics GPT flags as showing some form of moderation (Section~\ref{subsec:moderation-results}), BERT corroborates only \textit{animal rights} (exposure-moderated under both); it classifies \textit{death penalty}, \textit{progressive tax}, and \textit{smoking ban} as exposure-neutral instead.

\begin{table}[htbp]
    \centering
    \small
    \caption{\textbf{RQ1 labelling distribution across methods.} Category order matches Table~\ref{tab:classification-primary}. GPT total is 47 (national retail sales tax has no debates classified to it and is excluded); BERT total is 48.}
    \label{tab:primary-class-distribution}
    \begin{tabular}{lcccccc}
        \toprule
        Method & Conviction- & persuasiveness- & Jointly- & Conviction- & persuasiveness- & Indeter- \\
               & driven      & driven      & driven   & only        & only        & minate \\
        \midrule
        GPT-4o-mini        & 0 & 18 & 8 & 0 & 16 & 5 \\
        SBERT + BART-MNLI  & 0 & 22 & 4 & 0 & 21 & 1 \\
        \bottomrule
    \end{tabular}
\end{table}

\begin{table}[htbp]
    \centering
    \small
    \caption{\textbf{RQ2 labelling distribution across methods.} Category order matches Table~\ref{tab:classification-moderation}. GPT total is 47; BERT total is 48.}
    \label{tab:moderation-class-distribution}
    \begin{tabular}{lccc}
        \toprule
        Method & Exposure-moderated & Exposure-entrenched & Exposure-neutral \\
        \midrule
        GPT-4o-mini        & 3 & 1 & 43 \\
        SBERT + BART-MNLI  & 2 & 2 & 44 \\
        \bottomrule
    \end{tabular}
\end{table}

\clearpage
\subsection{Model comparison}\label{app:model-comparison}
Table~\ref{tab:model-comparison} summarizes the model comparison results based on expected log predictive density (ELPD) via leave-one-out cross-validation, computed with Pareto-smoothed importance sampling. Five specifications are compared: the full three-predictor model ($\conv+\expo+\pers$, ``regression''); the same model plus the $\conv \times \expo$ interaction (``interaction''); and three single-predictor models retaining only $\pers$, only $\conv$, or only $\expo$.

The regression model is the best-fitting specification (ELPD $=-25{,}700$), receiving 89\% of the stacking weight. The interaction model fits slightly worse ($\Delta\mathrm{ELPD} \approx -20$, $\Delta SE = 6.7$) and takes only 1\% of the weight, confirming that the $\conv \times \expo$ term does not earn its added complexity out-of-sample (Section~\ref{subsec:model-fit-interpr}). The persuasiveness-only model is a distant third ($\Delta\mathrm{ELPD} \approx -660$, weight 9\%), and the conviction-only ($\Delta\mathrm{ELPD} \approx -3{,}800$) and exposure-only ($\Delta\mathrm{ELPD} \approx -4{,}700$) models perform markedly worse still, with negligible weight. This ordering, full $\approx$ interaction $\gg$ persuasiveness-only $\gg$ conviction-only $>$ exposure-only, mirrors the coefficient-level finding in Section~\ref{subsec:primary-drivers} that persuasiveness alone captures most of the model's predictive power, with conviction adding a real but smaller increment and cross-cutting exposure adding very little on its own.

Three of the five models (regression, interaction, and exposure-only) are flagged by \texttt{arviz}~\cite{Martin2026arviz} for at least one unreliable Pareto-$k$ diagnostic ($k > 0.7$) among the $44{,}603$ held-out points. This flag is conservative and can trigger on a small number of points without compromising the ranking, so I inspect the pointwise diagnostics directly rather than treating the flag as disqualifying: for the regression model, only 2 of 44,603 points have $k$ in the unreliable range (max $k = 0.91$), for the interaction model, 5 points (max $k = 0.90$), and for the exposure-only model, 3 points (max $k = 0.73$); no model has any point with $k > 1$, the threshold at which PSIS-LOO estimates become altogether untrustworthy. The flagged points are therefore a small minority of a large evaluation set, and I read the ranking in Table~\ref{tab:model-comparison} as reliable.

\begin{table}[htbp]
\centering
\caption{\textbf{Model comparison based on expected log predictive density (ELPD).} Models are ranked by out-of-sample predictive performance, with higher ELPD indicating better fit. Reported quantities include the effective number of parameters (p), the difference in ELPD relative to the top-ranked model ($\Delta$ELPD), stacking weights, and associated standard errors (SE, $\Delta$SE). Warning flags a model with at least one Pareto-$k > 0.7$ point among the held-out set; see main text for the pointwise breakdown.}
\label{tab:model-comparison}
\begin{tblr}{
  width = \linewidth,
  colspec = {Q[220]Q[69]Q[110]Q[90]Q[110]Q[90]Q[69]Q[80]Q[113]},
  cells = {c},
  hline{1,2,7} = {-}{},
}
model                       & rank & ELPD     & p      & $\Delta$ELPD & weight & SE   & $\Delta$SE & warning \\
\textit{regression ($\conv{+}\expo{+}\pers$)} & 0    & -25700.0 & 1258.6 & 0.0        & 0.89   & 92.0 & 0.0        & True    \\
\textit{interaction ($+\conv{\times}\expo$)} & 1    & -25700.0 & 1304.8 & -20.0      & 0.01   & 92.0 & 6.7        & True    \\
\textit{persuasiveness-only ($\pers$)}   & 2    & -26300.0 & 1255.0 & -660.0     & 0.09   & 86.0 & 40.0       & False   \\
\textit{conviction-only ($\conv$)}    & 3    & -29500.0 & 1275.2 & -3800.0    & 0.02   & 53.0 & 81.0       & False   \\
\textit{exposure-only ($\expo$)}         & 4    & -30360.0 & 1383.1 & -4700.0    & 0.00   & 34.0 & 88.0       & True    
\end{tblr}
\end{table}
\clearpage
\subsection{Diversity-measure simulation}\label{app:diversity-simulation}
$\expo_{ij}$ (Eq.~\ref{eq:D}) is the negative mean pairwise cosine similarity among prior voters' belief vectors. This measure has one limitation: it cannot distinguish a large audience split into two internally homogeneous opposing camps from a large audience with genuinely dispersed, uncorrelated beliefs, since both converge toward the same value as audience size grows (same-camp and cross-camp pairs contribute oppositely to the average, and their counts scale together at the same rate). I demonstrate this directly with a controlled simulation.

I generate synthetic audiences of size $n \in \{2,3,5,8,12,20,35,50,75,100,150,200\}$ under three scenarios, each drawing $K=48$-dimensional belief vectors in $\{-1,0,+1\}^K$ (matching the paper's topic count), with 200 Monte Carlo trials per (scenario, $n$) pair: a \emph{homogeneous} audience clustered around one shared belief profile (a small per-voter flip probability and a base rate of ``no opinion'' entries added for realism); a \emph{polarized} audience split into two internally homogeneous, mutually opposed camps; and a \emph{diverse/uncorrelated} audience whose members' beliefs are drawn independently, with no shared structure at all.

\begin{figure}[htbp]
    \centering
    \includegraphics[width=0.7\linewidth]{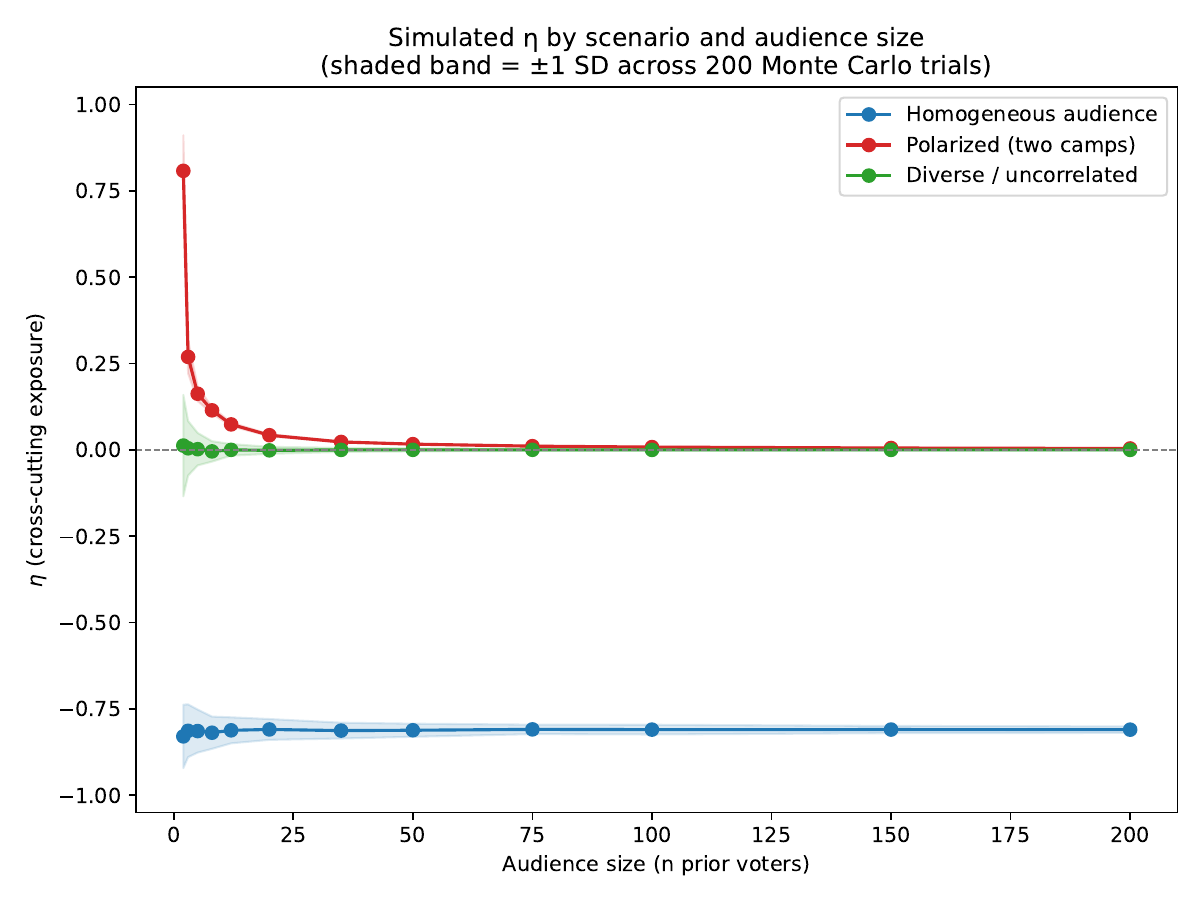}
    \caption{\textbf{Simulated $\expo$ by scenario and audience size.} Mean $\expo$ (Eq.~\ref{eq:D}) across 200 Monte Carlo trials per (scenario, $n$) pair, for three synthetic $K=48$-dimensional audience compositions: homogeneous, polarized (two internally homogeneous opposing camps), and diverse/uncorrelated. Shaded bands are $\pm 1$ SD across trials. The polarized scenario converges toward the diverse scenario's value as $n$ grows, becoming observationally indistinguishable from it by $n \approx 50$.}
    \label{fig:diversity-simulation}
\end{figure}

Figure~\ref{fig:diversity-simulation} shows the result. The homogeneous audience's $\expo$ stays stable at $\approx -0.81$ across all audience sizes, as expected. The diverse/uncorrelated audience's $\expo$ stays at $\approx 0$ throughout, from the smallest audiences onward. The polarized audience is the case of interest: at $n=2$, $\expo \approx 0.81$, correctly signalling a maximally split audience, but it falls sharply as $n$ grows -- $\expo \approx 0.12$ at $n=8$, $\expo \approx 0.02$ at $n=35$, and $\expo = 0.004$ at $n=200$, a value indistinguishable in practice from the diverse/uncorrelated audience's $\expo \approx 0.000$ at the same $n$. By $n \approx 50$ ($\expo=0.017$), the two scenarios are already effectively merged. A large, genuinely two-camp audience and a large, genuinely dispersed audience therefore become observationally near-identical under $\expo_{ij}$, even though they represent very different social environments.

As discussed in Section~\ref{sec:methods}, this limitation is less damaging to the moderation hypothesis than it might first appear, since cross-cutting exposure theory~\cite{mutz2006hearing} is concerned with exposure to disagreement per se, which both scenarios equally provide; it matters most for interpreting the single exposure-entrenched finding (smoking ban, Section~\ref{subsec:moderation-results}), since reactance and entrenchment accounts~\cite{bail2018exposure} more specifically implicate confrontation with a cohesive opposing camp, a distinction $\expo_{ij}$ cannot make once an audience is large.

\clearpage
\subsection{Cross-cutting exposure: temporal placebo check}\label{app:diversity-placebo}

An association between a debate's prior audience and a user's vote could be a genuine effect of that audience, or it could reflect users self-selecting into debates whose audience already resembles them, i.e., Manski's~\cite{manski1993} ``reflection problem,'' where endogenous and correlated effects are generally indistinguishable without further structure. Manski notes, however, that temporal structure can help: because $\expo_{ij}$ is built only from voters preceding user $i$, it is not mechanically confounded with $\conv_{ij}$, and it supports a placebo test based on timing.

I construct the placebo term $\tilde\expo_{ij}$ (Eq.~\ref{eq:D}, but computed over voters who cast their vote \emph{after} user $i$) and re-fit the single-predictor $\expo$-only model using $\tilde\expo_{ij}$ in place of $\expo_{ij}$, with the same standardization procedure applied independently to each. Because future votes cannot causally or compositionally precede a past vote, a real $\beta_\expo[t]$ that closely tracks its placebo counterpart $\tilde\beta_\expo[t]$ is consistent with $\beta_\expo[t]$ reflecting the ideological composition of a debate's audience as a static, time-invariant property, rather than a temporally-ordered association; a systematic difference between the two would instead indicate that timing matters, which selection into a static audience composition alone cannot easily explain.

\begin{figure}[htbp]
    \centering
    \includegraphics[width=\linewidth]{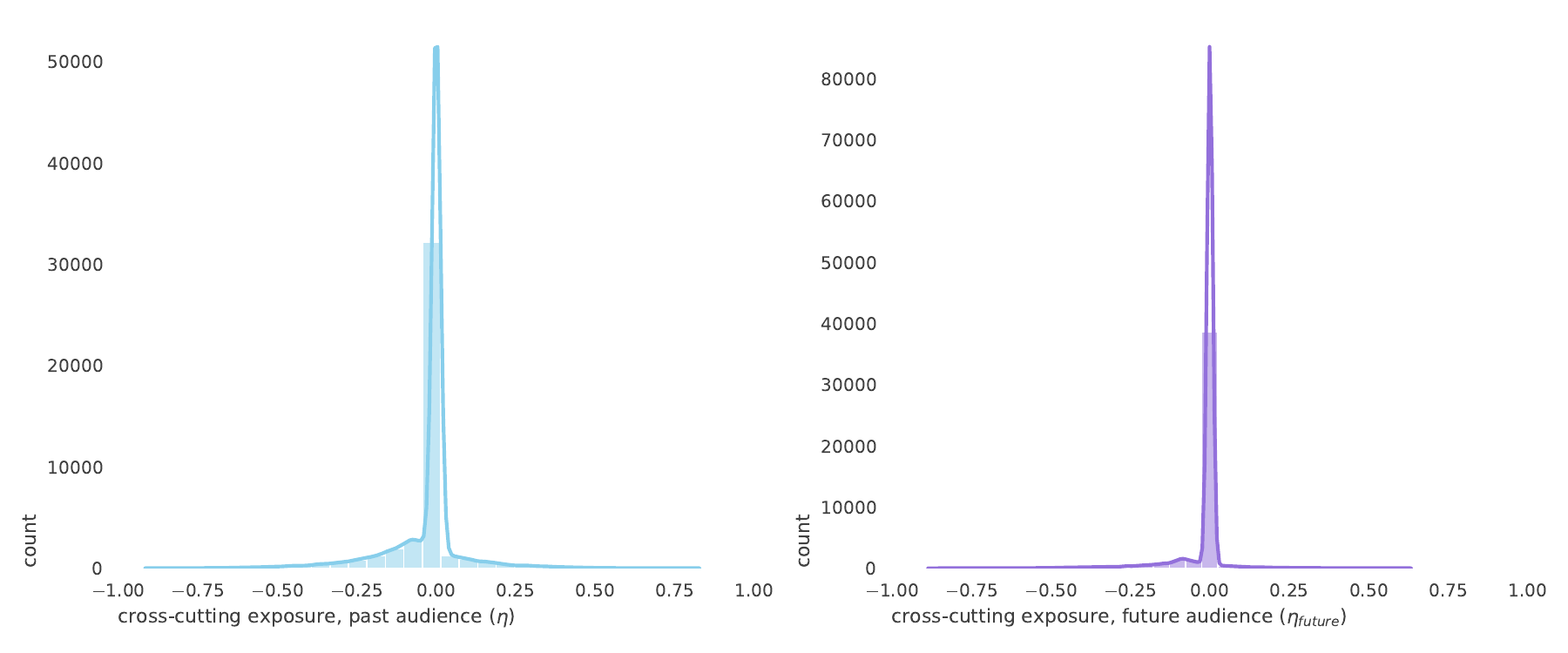}
    \caption{\textbf{Distribution of real (past-audience) vs.\ placebo (future-audience) cross-cutting exposure.} Histograms of $\expo_{ij}$ and $\tilde \expo_{ij}$ across all modelled votes, confirming comparable marginal distributions under the symmetric real/placebo construction.}
    \label{fig:D-placebo-distribution}
\end{figure}

\begin{figure}[htbp]
    \centering
    \includegraphics[width=0.55\linewidth]{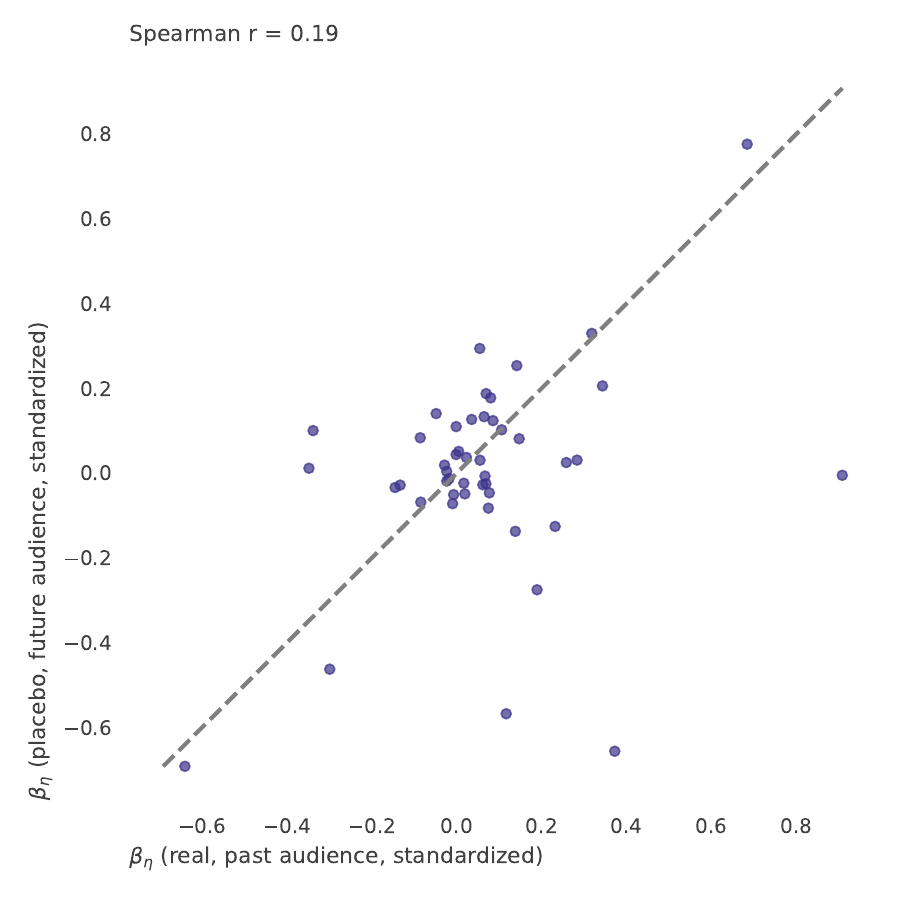}
    \caption{\textbf{Real vs.\ placebo $\beta_\expo$, by topic.} Each point is one topic's posterior mean $\beta_\expo[t]$ (real, past audience) against $\tilde\beta_\expo[t]$ (placebo, future audience), both from the standardized $\expo$-only model. Spearman $r=0.19$. The dashed line is the identity line.}
    \label{fig:D-placebo-scatter}
\end{figure}

Figure~\ref{fig:D-placebo-distribution} confirms that $\expo_{ij}$ and $\tilde\expo_{ij}$ have comparable marginal distributions, as expected given they are built from the same voter pool under a symmetric construction. Figure~\ref{fig:D-placebo-scatter} compares the two sets of topic-specific posterior means directly: the raw (Spearman) correlation across all 47 topics is $r=0.19$, which is low, but is consistent with both quantities being weakly identified at the topic level.

The more direct test is the paired posterior difference $\beta_\expo[t] - \tilde\beta_\expo[t]$, computed per topic from matched posterior draws. Using each topic's raw 90\% credible interval, 7 of 47 topics show a difference that excludes zero: abortion, capitalism, environmental protection, free trade, national health care, united nations, and welfare.

\subsection{The no-prior-audience case}\label{app:no-prior}
Votes with no observed prior audience are modelled with a single global intercept $\gamma_{\text{no-prior}}$ in place of the two $\expo$-pathway terms. A global treatment is appropriate because the no-prior-audience case reflects a structural difference in the voting situation itself, e.g.\ debaters trivially voting their declared side, which is expected to hold uniformly across topics; a per-topic $\gamma$ would in any case be poorly identified. In the fitted regression model, $\hat\gamma_{\text{no-prior}} = -0.043$ (90\% CI $[-0.068, -0.019]$), i.e.\ such votes are credibly \emph{less} likely to be \textbf{PRO} than the $\expo$-pathway alone would predict.
%TC:endignore
\end{document}